\pdfoutput=1
\documentclass[aps,prd,amsmath,floats,floatfix, twocolumn,
superscriptaddress,nofootinbib,showpacs,longbibliography]{revtex4-1}
\usepackage[T1]{fontenc}
\usepackage[utf8]{inputenc}
\usepackage{txfonts}
\usepackage{verbatim}
\usepackage[dvipsnames, usenames]{xcolor}
\definecolor{linkcolor}{rgb}{0.0,0.3,0.5}
\usepackage[hypertexnames=false, unicode, colorlinks=true, linkcolor=linkcolor,
citecolor=linkcolor, filecolor=linkcolor,urlcolor=linkcolor,
pdfusetitle]{hyperref}
\usepackage[all]{hypcap}
\usepackage{graphicx}
\usepackage{xspace}
\usepackage{amssymb}
\usepackage{microtype}
\usepackage{subfigure}

\usepackage{url}

\usepackage[english]{babel}
\usepackage{blindtext}

\usepackage[normalem]{ulem} 
\usepackage{bm} 
\usepackage{mathrsfs}

\DeclareMathAlphabet{\mathpzc}{OT1}{pzc}{m}{it}

\usepackage[nomessages]{fp}

\begin{document}
\title{Adding higher-order spherical harmonics in non-spinning eccentric \\binary black hole merger waveform models}
\newcommand{\KITP}{\affiliation{Kavli Institute for Theoretical Physics, University of California Santa Barbara, Kohn Hall, Lagoon Rd, Santa Barbara, CA 93106}}
\newcommand{\UMassDMath}{\affiliation{Department of Mathematics,
		University of Massachusetts, Dartmouth, MA 02747, USA}}
\newcommand{\UMassDPhy}{\affiliation{Department of Physics,
		University of Massachusetts, Dartmouth, MA 02747, USA}}
\newcommand{\CSCVRUMass}{\affiliation{Center for Scientific Computing and Data Science Research, University of Massachusetts, Dartmouth, MA 02747, USA}}
\newcommand{\URI}{\affiliation{Department of Physics and Center for Computational Research, University of Rhode Island, Kingston, RI 02881, USA}}   
\newcommand{\TAPIR}{\affiliation{Theoretical AstroPhysics Including Relativity and Cosmology, California Institute of Technology, Pasadena, California, USA}}
\newcommand{\AEI}{\affiliation{Max Planck Institute for Gravitational Physics (Albert Einstein Institute), Am Mühlenberg 1, Potsdam 14476, Germany}}

\author{Tousif Islam}
\email{tislam@kitp.ucsb.edu}
\KITP
\TAPIR

\author{Gaurav Khanna}
\URI
\UMassDPhy
\CSCVRUMass

\author{Scott E. Field}
\UMassDMath
\CSCVRUMass

\hypersetup{pdfauthor={Islam et al.}}

\date{\today}

\begin{abstract}
\texttt{gwNRHME} is a recently developed framework that seamlessly converts a multi-modal (i.e with several spherical harmonic modes) quasi-circular waveform into multi-modal eccentric waveform if the quadrupolar eccentric waveform is known. Here, we employ the \texttt{gwNRHME} framework to combine a multi-modal quasi-circular numerical relativity surrogate waveform model \texttt{NRHybSur3dq8} and quadrupolar non-spinning post-Newtonian eccentric waveform model \texttt{EccentricIMR} to construct multi-modal non-spinning eccentric model \texttt{NRHybSur3dq8-gwNRHME}. Using a total of 35 eccentric numerical relativity (NR) simulations obtained from the SXS and RIT catalogs, we demonstrate that \texttt{NRHybSur3dq8-gwNRHME} model predictions agree well with NR (with typical relative $L_2$ errors of $\sim 0.01$ for the dominant quadrupolar mode) for mass ratios $ 1 \leq q \leq 4$ and eccentricities up to $\sim 0.2$ measured about 10 cycles before the merger. Our frequency-domain mismatches (calculated assuming advanced LIGO design sensitivity curve) are mostly below 0.01. To demonstrate the modularity of the \texttt{gwNRHME} framework, we further combine \texttt{EccentricIMR} with \texttt{BHPTNRSur1dq1e4} model and develop a non-spinning eccentric models named \texttt{BHPTNRSur1dq1e4-gwNRHME}. Finally, we develop a different variant of these models by replacing \texttt{EccentricIMR} with \texttt{EccentricTD}. Both the \texttt{gwNRHME} framework and associated models are available through the \texttt{gwModels} package.
\end{abstract}
\maketitle
\section{Introduction}
Gravitational waves (GWs) emitted from binary black hole (BBH) mergers can be 
expressed as a superposition of $-2$ spin-weighted spherical harmonic modes with indices $(\ell,m$):
\begin{align}
h(t,\theta,\phi;\boldsymbol{\lambda}) &= \sum_{\ell=2}^\infty \sum_{m=-\ell}^{\ell} h_{\ell m}(t;\boldsymbol\lambda) \; _{-2}Y_{\ell m}(\theta,\phi)\,,
\label{hmodes}
\end{align}
where $\boldsymbol{\lambda}$ describes the masses, spins and eccentricities of the binary, and ($\theta$,$\phi$) are angles describing the orientation of the binary.
Building an accurate model for the waveform
is essential for faithful and efficient source characterization. While such models exist for non-spinning and generically spinning quasi-circular BBH mergers, incorporating eccentricity into those models is still in its nascent stage~\cite{Tiwari:2019jtz, Huerta:2014eca, Moore:2016qxz, Damour:2004bz, Konigsdorffer:2006zt, Memmesheimer:2004cv,Hinder:2017sxy, Cho:2021oai,Chattaraj:2022tay,Hinderer:2017jcs,Cao:2017ndf,Chiaramello:2020ehz,Albanesi:2023bgi,Albanesi:2022xge,Chiaramello:2020ehz,Ramos-Buades:2021adz,Liu:2023ldr,Huerta:2016rwp,Huerta:2017kez,Joshi:2022ocr,Setyawati:2021gom,Wang:2023ueg,Islam:2021mha,Carullo:2023kvj,Nagar:2021gss,Khalil:2021txt,Gamba:2024cvy}. The unavailability of faithful and computationally cheap multi-modal waveform models for eccentric BBH mergers is currently a potential bottleneck in conclusively identifying and characterizing such mergers in the recorded LIGO-Virgo-KAGRA (LVK) strain data~\cite{Harry:2010zz,VIRGO:2014yos,KAGRA:2020tym,LIGOScientific:2018mvr,LIGOScientific:2020ibl,LIGOScientific:2021usb,LIGOScientific:2021djp}, even if these signals are present. Eccentric binaries, however, are among the most interesting sources observable in current-generation detectors. These binaries are expected to form in dense globular clusters and galactic nuclei, making them a valuable probe to study the environmental properties of these clusters/nuclei. On a population level, binaries formed in dense globular clusters and binaries formed in isolated environments are expected to exhibit different eccentricity distributions~\cite{Rodriguez:2017pec,Rodriguez:2018pss,Samsing:2017xmd,Zevin:2018kzq,Zevin:2021rtf}. Observing tens of eccentric BBH mergers can, therefore, help us understand the formation channels for BBHs in more detail. Furthermore, the detection of eccentric BBH mergers could provide new opportunities to test the general theory of relativity in strong field regime~\cite{Ma:2019rei}. While there are some possible hints of eccentricity in detected LVK strain data~\cite{Romero-Shaw:2020thy,Gayathri:2020coq,Gamba:2021gap,Ramos-Buades:2023yhy,Gupte:2024jfe}, no conclusive evidence has been found yet.

\begin{figure}
\includegraphics[width=\columnwidth]{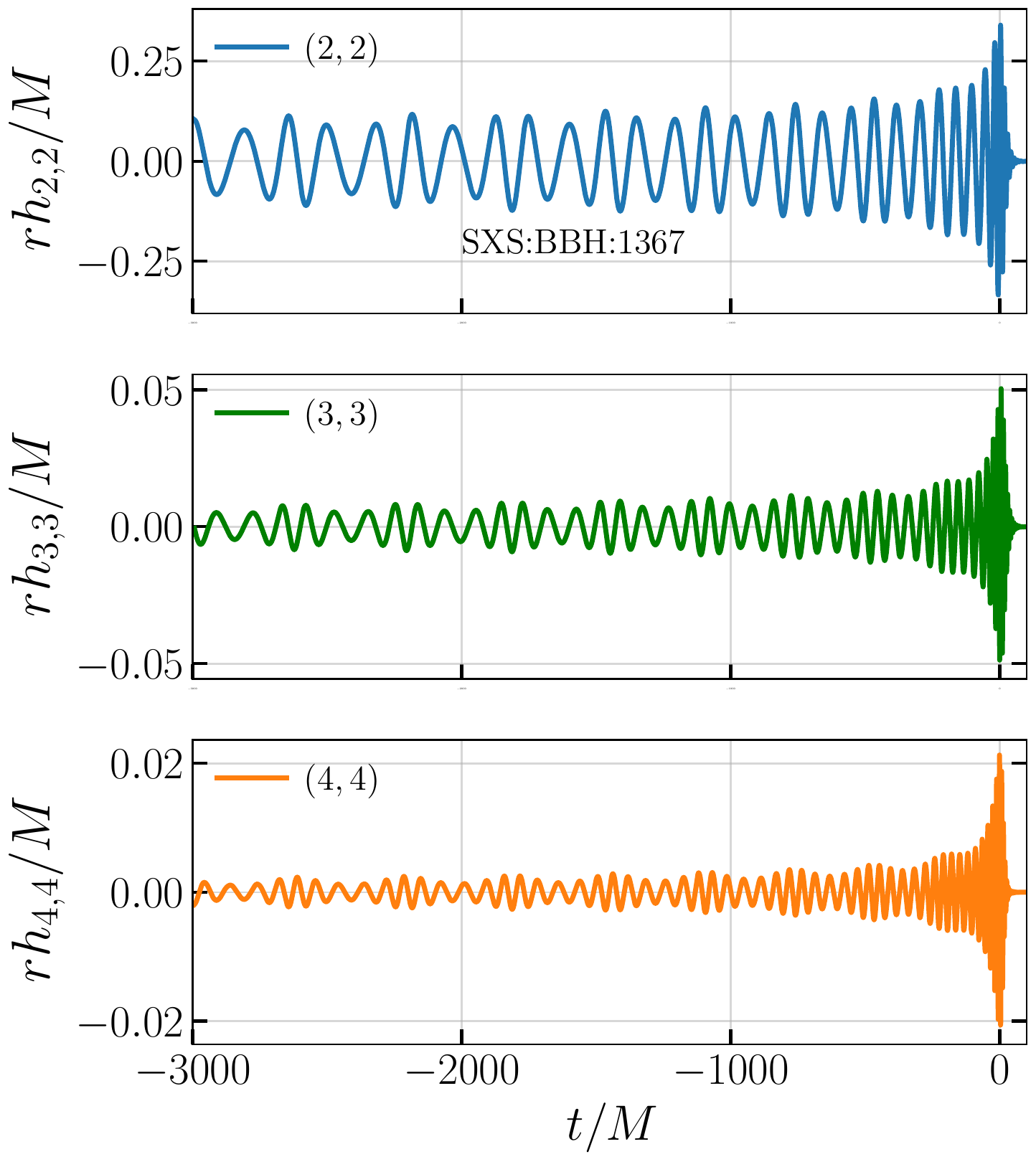}
\caption{We show the real parts of the $(2,2)$ (upper panel; blue line), $(3,3)$ (middle panel; green line) and $(4,4)$ (lower panel; orange line) spherical harmonic modes of gravitational waveform extracted from an non-spinning eccentric BBH simulation \texttt{SXS:BBH:1367}. This simulation is characterized by mass ratio $q=2$, eccentricity $e_{\rm ref}=0.1$ and mean anomaly $l_{\rm ref}=-0.743$ measured at a dimensionless reference frequency of $x=0.75$~\cite{Hinder:2017sxy}.}
\label{fig:SXS1367_waveforms}
\end{figure}

Currently available waveform models for eccentric BBH mergers employ various frameworks and techniques, including post-Newtonian (PN) approximations~\cite{Huerta:2016rwp,Huerta:2017kez,Joshi:2022ocr}, the effective-one-body formalism~\cite{Hinderer:2017jcs,Cao:2017ndf,Chiaramello:2020ehz,Albanesi:2023bgi,Albanesi:2022xge,Riemenschneider:2021ppj,Chiaramello:2020ehz,Ramos-Buades:2021adz,Liu:2023ldr,Nagar:2021gss}, semi-analytical modeling based on numerical relativity (NR) simulations~\cite{Setyawati:2021gom,Wang:2023ueg,Carullo:2023kvj,Hinder:2017sxy}, and data-driven methods~\cite{Islam:2021mha}. Some of these models are computationally expensive, and not all models extend to the merger phase. Among these, only a handful of models (\texttt{NRSur2dq1Ecc}~\cite{Islam:2021mha}, \texttt{SEOBNRE}~\cite{Cao:2017ndf,Liu:2023ldr}, \texttt{SEOBNRv4EHM}~\cite{Ramos-Buades:2021adz}, \texttt{TEORBResumS}~\cite{Chiaramello:2020ehz,Nagar:2021gss}) include subdominant spherical harmonics modes in addition to the dominant quadrupolar mode. Recently, we have empirically identified a simple, mode-independent relation between circular and eccentric non-spinning BBH merger waveforms using publicly available NR data~\cite{Islam:2024rhm}. These relations, in turn, 
can be leveraged to convert a quasi-circular multi-modal waveform model into an eccentric multi-modal waveform model if a quadrupolar eccentric model is available~\cite{Islam:2024rhm}. We shall refer to this framework as \texttt{gwNRHME}, and eccentric, multi-modal GW models that are built from it will be named \texttt{X-gwNRHME}, where \texttt{X} is the non-eccentric ``carrier'' model whose harmonic modes we shall modify according to the \texttt{gwNRHME} prescription.
Models built from this framework is publicly hosted at the \texttt{gwModels} Python package and can be accessed at \href{https://github.com/tousifislam/gwModels}{https://github.com/tousifislam/gwModels}.

In this paper, we demonstrate how the \texttt{gwNRHME} framework can be used to build multi-modal eccentric non-spinning waveform models by combining current quadrupolar non-spinning eccentric models with circular multi-modal waveform models. We choose one of the publicly available eccentric inspiral-merger-ringdown models named \texttt{EccentricIMR}~\cite{Hinder:2008kv}. This model is developed by combining a PN inspiral waveform model with a quasi-circular merger waveform model. The inspiral part of the waveform includes contributions up to 3PN order conservative and 2PN order reactive terms to the BBH dynamics~\cite{Hinder:2008kv}. While the original model is implemented in \texttt{Mathematica}, we call it through a Python wrapper available through the \texttt{gwModels} package. Additionally, we incorporate \texttt{EccentricTD}~\cite{Tanay:2016zog}, 
a recent inspiral-only eccentric non-spinning waveform models. For circular waveform generation, we employ the NR surrogate model named \texttt{NRHybSur3dq8}~\cite{Varma:2018mmi} as our default choice, along with \texttt{BHPTNRSur1dq1e4}~\cite{Islam:2022laz} and \texttt{IMRPhenomTHM}~\cite{Estelles:2020twz} models.

The rest of the paper is organized as follows. We describe the phenomenology of eccentric, non-spinning waveforms obtained from NR and the basics of the \texttt{gwNRHME} framework in Section~\ref{sec:phenomenology}. We then describe the methods followed to extend \texttt{EccentricIMR} and \texttt{EccentricTD} models to include higher order spherical harmonic modes in Section~\ref{sec:add_HMs}. We discuss our results, point out model limitations, and outline future directions in Section~\ref{sec:conclusion}.

\begin{figure}
\includegraphics[width=\columnwidth]{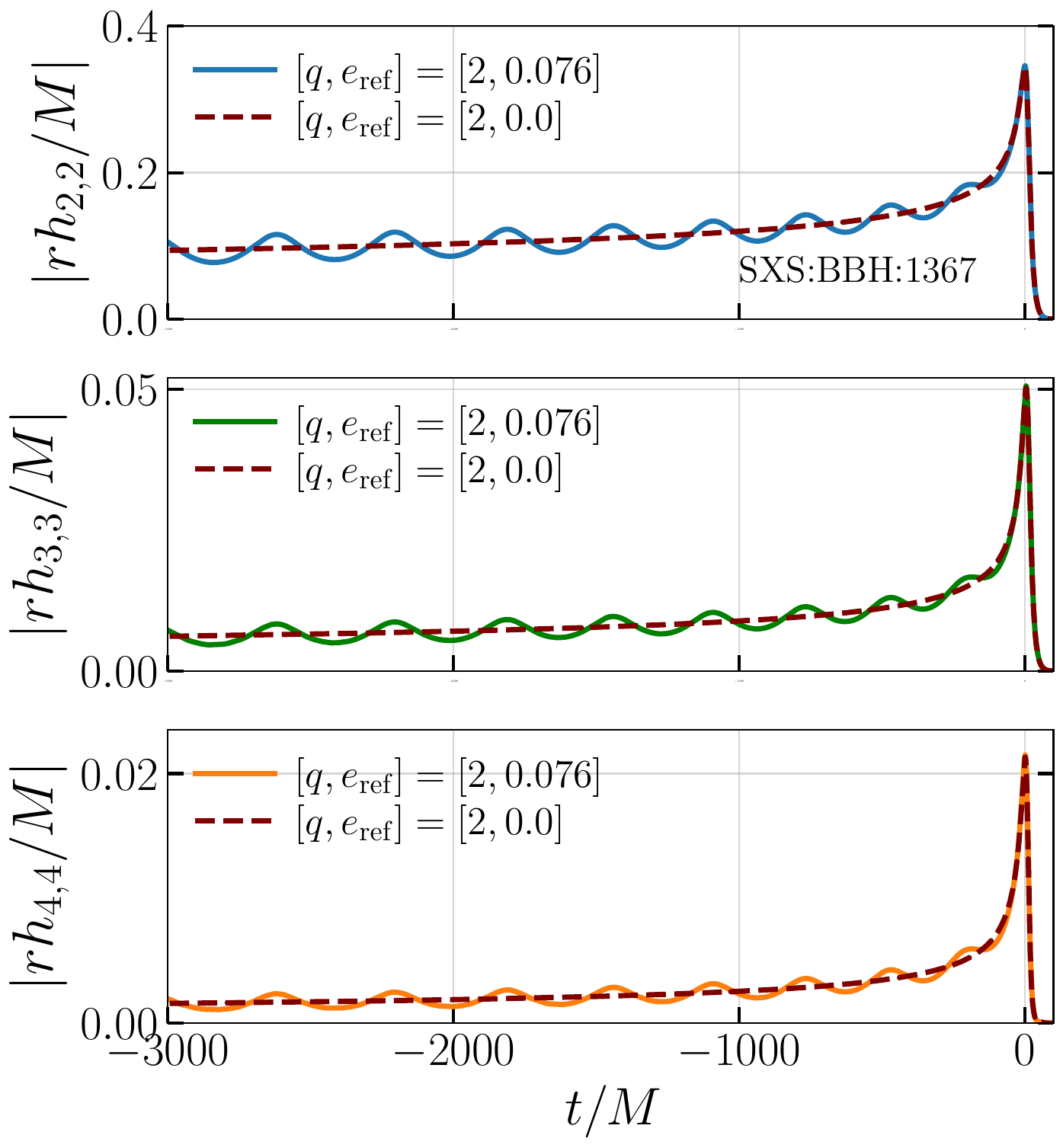}
\caption{We show the amplitudes of the $(2,2)$ (upper panel; blue line), $(3,3)$ (middle panel; green line) and $(4,4)$ (lower panel; orange line) spherical harmonic modes of gravitational waveform extracted from an non-spinning eccentric BBH simulation \texttt{SXS:BBH:1367}. This simulation is characterized by mass ratio $q=2$, eccentricity $e_{\rm ref}=0.1$ and mean anomaly $l_{\rm ref}=-0.743$ measured at a dimensionless reference frequency of $x=0.75$~\cite{Hinder:2017sxy}. In addition, we show the corresponding circular amplitudes obtained from the \texttt{SXS:BBH:0184} simulation as dashed maroon lines.}
\label{fig:SXS1367_amplitude}
\end{figure}

\begin{figure}
\includegraphics[width=\columnwidth]{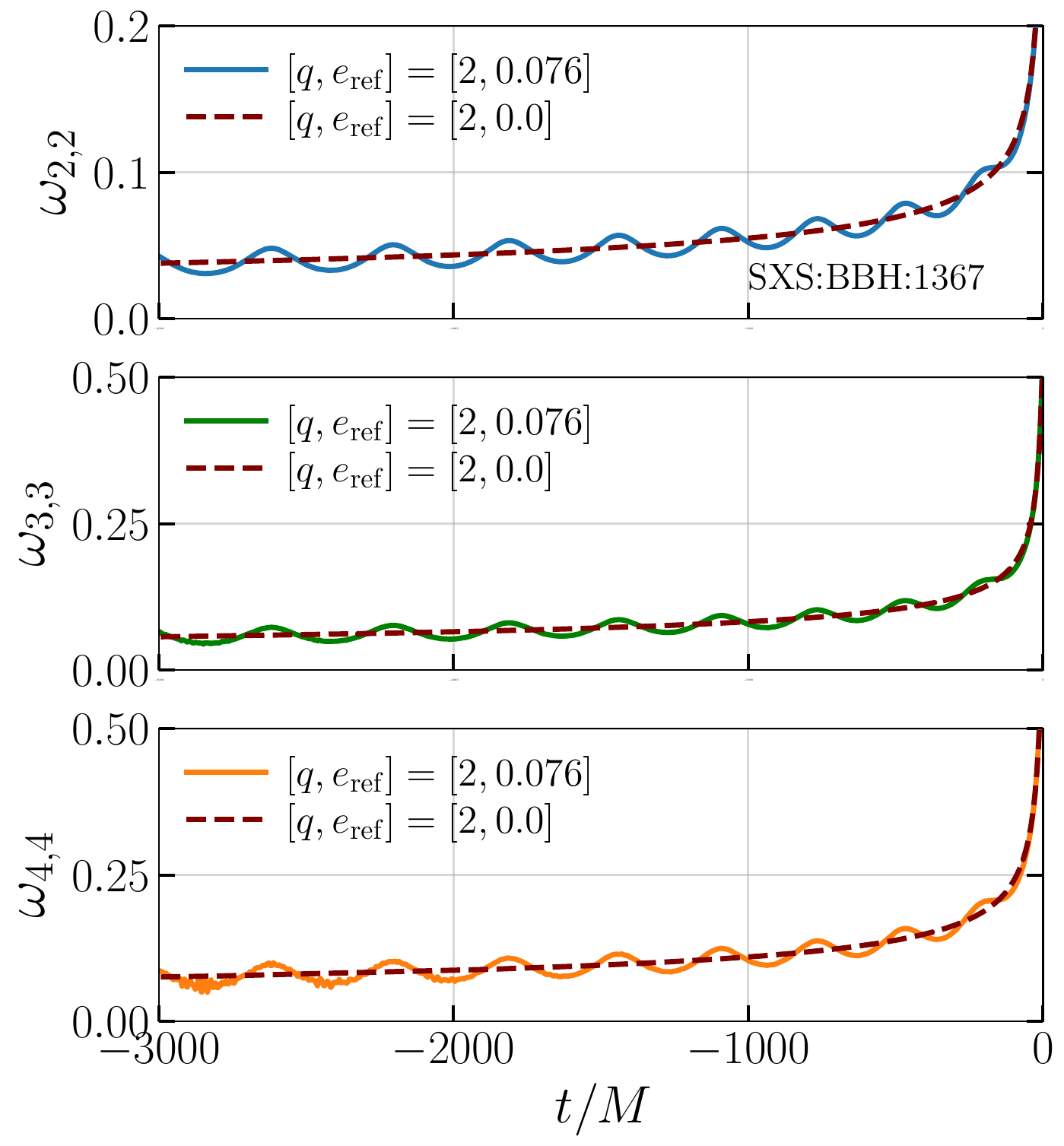}
\caption{We show the instantaneous frequencies of the $(2,2)$ (upper panel; blue line), $(3,3)$ (middle panel; green line) and $(4,4)$ (lower panel; orange line) spherical harmonic modes of gravitational waveform extracted from an non-spinning eccentric BBH simulation \texttt{SXS:BBH:1367}. This simulation is characterized by mass ratio $q=2$, eccentricity $e_{\rm ref}=0.1$ and mean anomaly $l_{\rm ref}=-0.743$ measured at a dimensionless reference frequency of $x=0.75$~\cite{Hinder:2017sxy}. In addition, we show the corresponding frequencies in the circular case, obtained from the \texttt{SXS:BBH:0184} simulation, as dashed maroon lines.}
\label{fig:SXS1367_frequency}
\end{figure}

\section{Phenomenology of eccentric BBH merger waveforms}
\label{sec:phenomenology}
We express the masses (and times) in geometric units, i.e., we consider $G=c=1$, and all binaries are scaled to have a total mass of $M:=m_1+m_2=1$ (where $m_1$ and $m_2$ are the masses of the larger and smaller black holes respectively). As we restrict ourselves to non-spinning eccentric binaries, we have $\boldsymbol{\lambda}:=\{q,e_{\rm ref},l_{\rm ref}\}$.
Here, $e_{\rm ref}$ is the eccentricity and $l_{\rm ref}$ is the mean anomaly estimated at a chosen reference time or frequency. There could be multiple ways to define eccentricity and the choice of eccentricity estimator will not change the results.
We decompose each complex-valued spherical harmonic mode $h_{\ell m}(t; \boldsymbol\lambda)$ into a real-valued amplitude $A_{\ell m}(t)$ and phase $\phi_{\ell m}(t)$ such that:
\begin{equation}
h_{\ell m}(t;q,e_{\rm ref}) = A_{\ell m}(t) e^{i \phi_{\ell m}(t)}.
\label{eq:amp_phase}
\end{equation}
The instantaneous frequency of each spherical harmonic mode is then obtained as:
\begin{equation}
\omega_{\ell m}(t;q,e_{\rm ref}) = \frac{d\phi_{\ell m}(t)}{dt}.
\label{eq:freq}
\end{equation}
Orbital angular frequency of the binary is then: $\omega_{\text{orb}}=0.5 \times \omega_{22}$. We define the time coordinate such that the maximum amplitude of the $(\ell,m)=(2,2)$ mode occurs at $t=0$. Here, we only focus on the positive $m$ modes as the negative $m$ modes are obtained by the symmetry relation: $h_{\ell m} = (-1)^{\ell} h^{*}_{\ell-m}$ where $*$ indicates complex conjugate.

\subsection{NR data}
We utilize a total of 15 publicly available eccentric NR simulations from the SXS catalog and an additional 20 eccentric simulations from the RIT catalog. SXS NR data has mass ratios ranging from $q=1$ to $q=3$. The eccentricity and mean anomaly of the SXS NR simulations are measured at a dimensionless frequency of $x=(M \omega_{\text{orb}})^{2/3}$~\cite{Hinder:2008kv} and reaches up to $e_{\rm ref}=0.2$. The RIT simulations also exhibit similar ranges of eccentricity (as measured at the start of the waveform), with mass ratios spanning between $q=1$ and $q=4$.
In Figure~\ref{fig:SXS1367_waveforms}, we show the real parts of the $(2,2)$ (upper panel; blue line), $(3,3)$ (middle panel; green line) and $(4,4)$ (lower panel; orange line) spherical harmonic modes of gravitational waveform extracted from an non-spinning eccentric BBH simulation \texttt{SXS:BBH:1367}. This simulation is characterized by mass ratio $q=2$, eccentricity $e_{\rm ref}=0.1$ and mean anomaly $l_{\rm ref}=-0.743$ measured at a dimensionless reference frequency of $x=0.75$~\cite{Hinder:2017sxy}. We then show the amplitudes and instantaneous frequencies of these modes in Figures~\ref{fig:SXS1367_amplitude} and \ref{fig:SXS1367_frequency} respectively. For comparison, we also show the amplitudes and instantaneous frequencies in the circular case (\texttt{SXS:BBH:0184}). It turns out that the circular amplitude and instantaneous frequencies go right through the eccentric amplitudes and frequencies.

\begin{figure}
\includegraphics[width=\columnwidth]{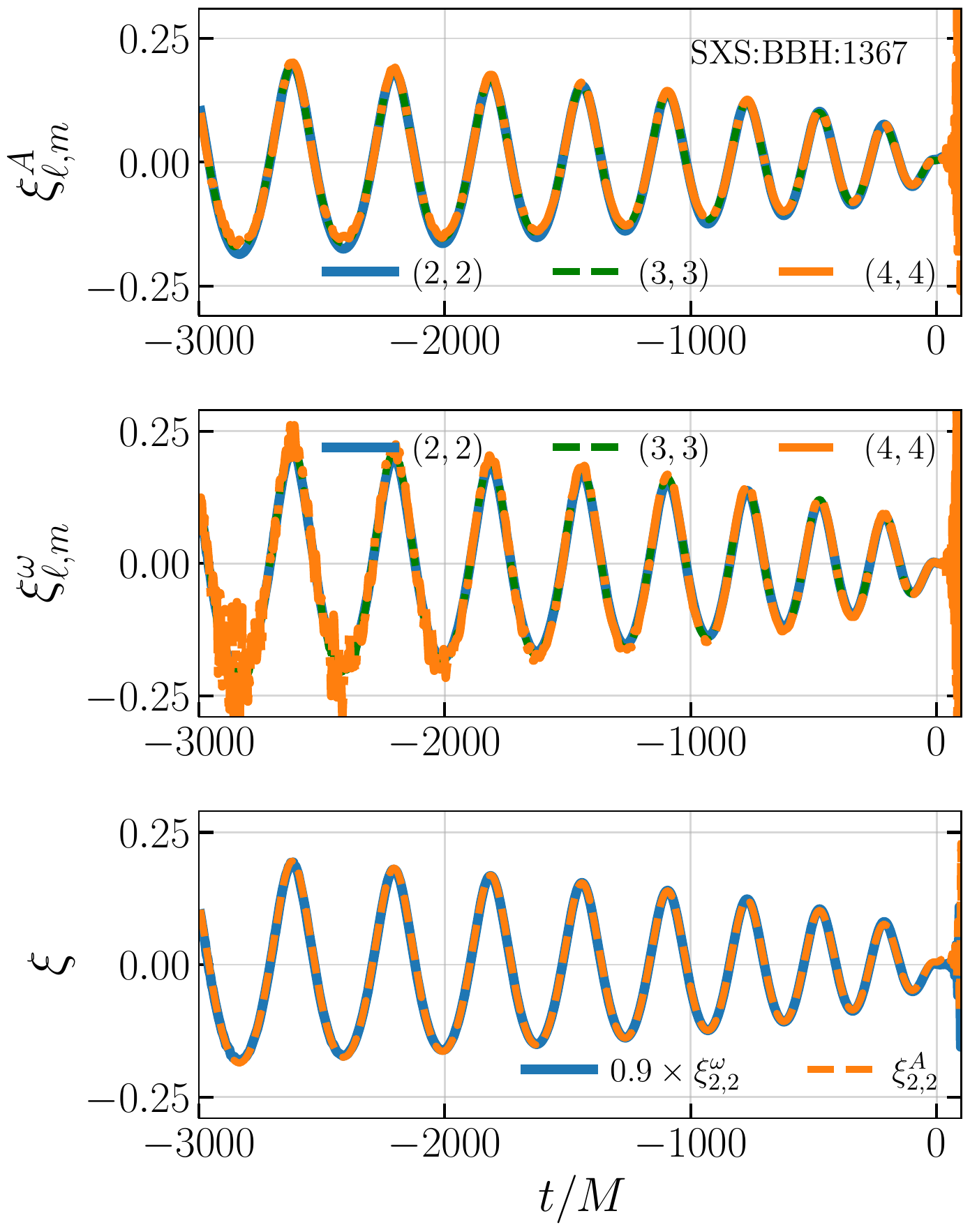}
\caption{We show the eccentric modulations in amplitudes $\xi_{\ell,m}^{A}$ (upper panel) and in frequencies $\xi_{\ell, m}^{\omega}$ (middle panel) for three representative modes: $(2,2)$ (blue) $(3,3)$ (green) and $(4,4)$ (orange) for a binary with mass ratio $q=2$, eccentricity $e_{\rm ref}=0.1$ and mean anomaly $l_{\rm ref}=-0.743$ measured at a reference dimensionless frequency of $x_{\text{ref}}=0.075$. We extract this modulations from the eccentric NR simulation \texttt{SXS:BBH:1367} and the corresponding circular simulation \texttt{SXS:BBH:0184}~\cite{Hinder:2008kv}. In the lower panel, we demonstrate that these two modulations are related by a factor of $K=0.9$ (obtained through a phenomenological fit provided in Ref.~\cite{Islam:2024rhm}).}
\label{fig:SXS1367_modulations}
\end{figure}

\subsection{Eccentric modulations}
Following Ref.~\cite{Islam:2024rhm}, we then compute the eccentric amplitude modulations and eccentric frequency modulations for different NR simulations. Eccentric frequency modulation for each mode is given as:
\begin{equation}
\xi_{\ell m}^{\omega}(t;q,e_{\rm ref},l_{\rm ref}) = b_{\ell m}^\omega \frac{\omega_{\ell m}(t;q,e_{\rm ref},l_{\rm ref})-\omega_{\ell m}(t;q,e_{\rm ref}=0)}{\omega_{\ell m}(t;q,e_{\rm ref}=0)}.
\label{eq:freq_mod}
\end{equation}
Amplitude modulations is written as:
\begin{equation}
\xi_{\ell m}^{A}(t;q,e_{\rm ref},l_{\rm ref}) = b^{A}_{\ell m} \frac{2}{\ell} \frac{A_{\ell m}(t;q,e_{\rm ref},l_{\rm ref})-A_{\ell m}(t;q,e_{\rm ref}=0)}{A_{\ell m}(t;q,e_{\rm ref}=0)}.
\label{eq:amp_mod}
\end{equation}
Here, $\xi_{\ell m}^{\omega}(t;q,e_{\rm ref},l_{\rm ref})$ does not have any mode dependence while $\xi_{\ell m}^{A}(t;q,e_{\rm ref},l_{\rm ref})$ depends on the $\ell$ value of the spherical harmonic mode. We set the constants $b_{\ell m}^\omega$ and $b^{A}_{\ell m}$ to be unity. 
Furthermore, the amplitude modulations and frequency modulations are related by a scaling factor $B=0.9$ (obtained through phenomenological fits~\cite{Islam:2024rhm}) such that
\begin{equation}
\label{eq:amp_freq_mod_relation}
\xi_{\ell m}^{A}(t;q,e_{\rm ref},l_{\rm ref}) = B \xi_{\ell m}^{\omega}(t;q,e_{\rm ref},l_{\rm ref}).
\end{equation}
We demonstrate this universal feature of the eccentric modulations for \texttt{SXS:BBH:1367}. Figure~\ref{fig:SXS1367_modulations} clearly shows that the amplitude modulations obtained from different spherical harmonic modes are the same (upper panel). The frequency modulations in different spherical harmonic modes are also the same (middle panel). Finally, amplitude and frequency modulations are related to each other by the factor $B$ (lower panel).

\subsection{Overview of \texttt{gwNRHME} model}
\label{sec:gwNRHME}
Based on these eccentric modulations, the \texttt{gwNRHME} framework (available through the \texttt{gwModels} package~\cite{Islam:2024rhm}) is developed, which combines a multi-modal quasi-circular waveform model $h^{\rm Cir}_{\ell,m}(t; q, e_{\rm ref}=0)$ with a quadrupolar eccentric waveform model $h^{\rm Ecc}_{2,2}(t; q, e_{\rm ref},l_{\rm ref})$ to provide a multi-modal eccentric waveform model $h^{\rm Ecc}_{\ell,m}(t; q, e_{\rm ref},l_{\rm ref})$. 
While detailed information about \texttt{gwNRHME} framework is provided in the Ref.~\cite{Islam:2024rhm}, here, we outline the steps briefly.

1. First, we select \texttt{NRHybSur3dq8}~\cite{Varma:2018mmi} as our base quasi-circular aligned-spin waveform model. This model is trained based on 104 NR waveforms with mass ratios $q \leq 8$ and spins $|\chi_{1,2}| \leq 0.8$. It encompasses the $\ell \leq 4$ and $(5,5)$ spin-weighted spherical harmonic modes but excludes the $(4,1)$ or $(4,0)$ modes. One can however replace \texttt{NRHybSur3dq8} with any other quasi-circular model of choice.

2. Next, we employ one of the quadrupolar eccentric waveform models. In this paper, we utilize the \texttt{EccentricIMR} model (in Section~\ref{sec:eccentricimr})~\cite{Hinder:2017sxy} and \texttt{EccentricTD} model~\cite{Tanay:2016zog} (in Section~\ref{sec:eccentrictd}) as our eccentric base models.

\begin{figure*}
\includegraphics[width=\textwidth]{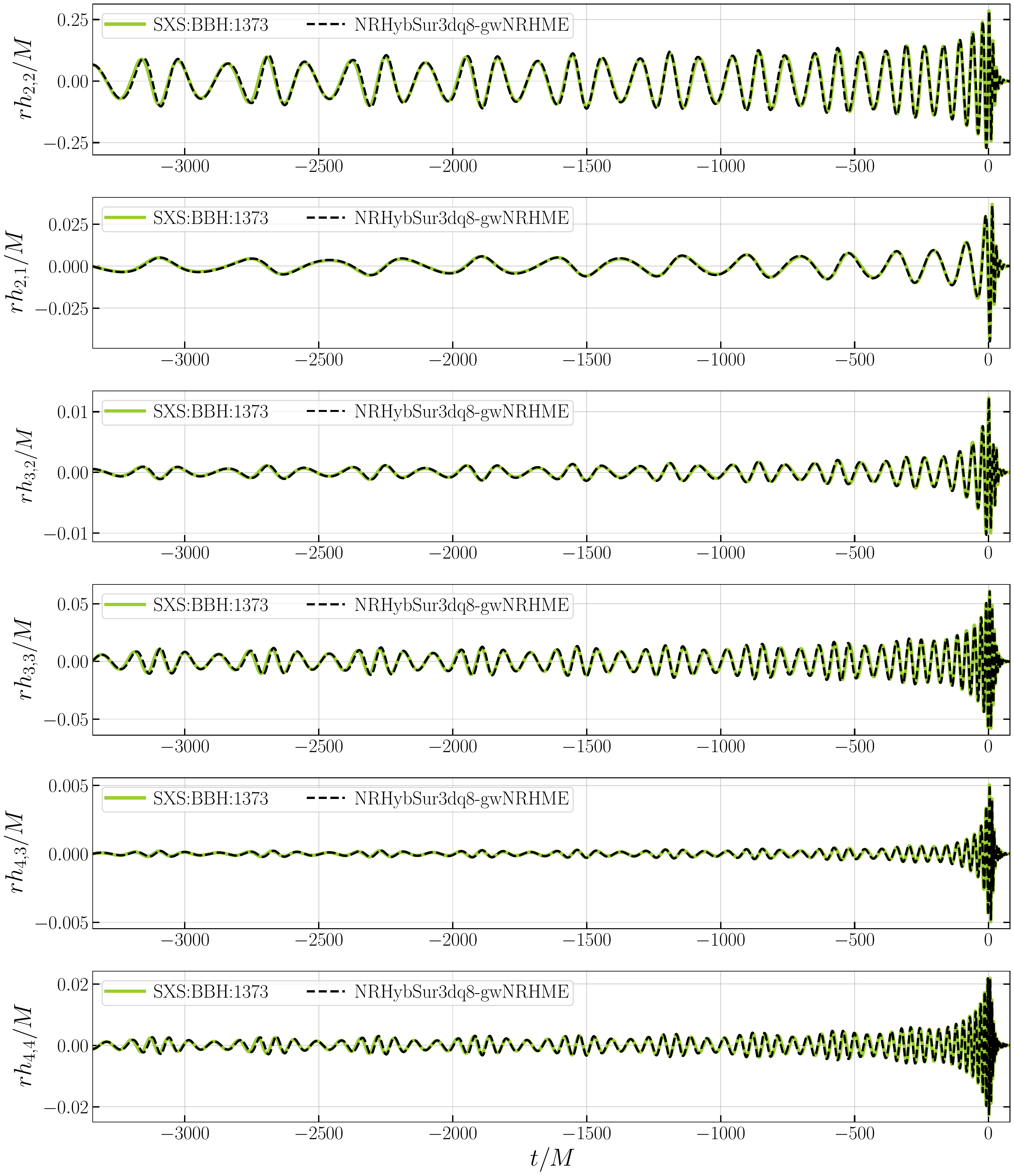}
\caption{We show the eccentric spherical harmonic modes (black dashed lines) obtained 
from the \texttt{NRHybSur3dq8-gwNRHME} model and corresponding NR data from \texttt{SXS:BBH:1373} simulation (green solid lines). We obtain \texttt{NRHybSur3dq8-gwNRHME} predictions by combining quadrupolar eccentric model \texttt{EccentricIMR} and circular waveform model \texttt{NRHybSur3dq8} using \texttt{gwNRHME} framework (available at \href{https://github.com/tousifislam/gwModels}{https://github.com/tousifislam/gwModels}). This simulation is characterized by mass ratio $q=2$ and eccentricity $e_{\rm ref}=0.09$ measured at a reference dimensionless frequency $x_{\rm ref}=0.075$. We find that \texttt{NRHybSur3dq8-gwNRHME} predictions are visually indistinguishable from NR.}
\label{fig:SXS1371_NRHMEcc_wfs}
\end{figure*}

\begin{figure*}
\includegraphics[width=\textwidth]{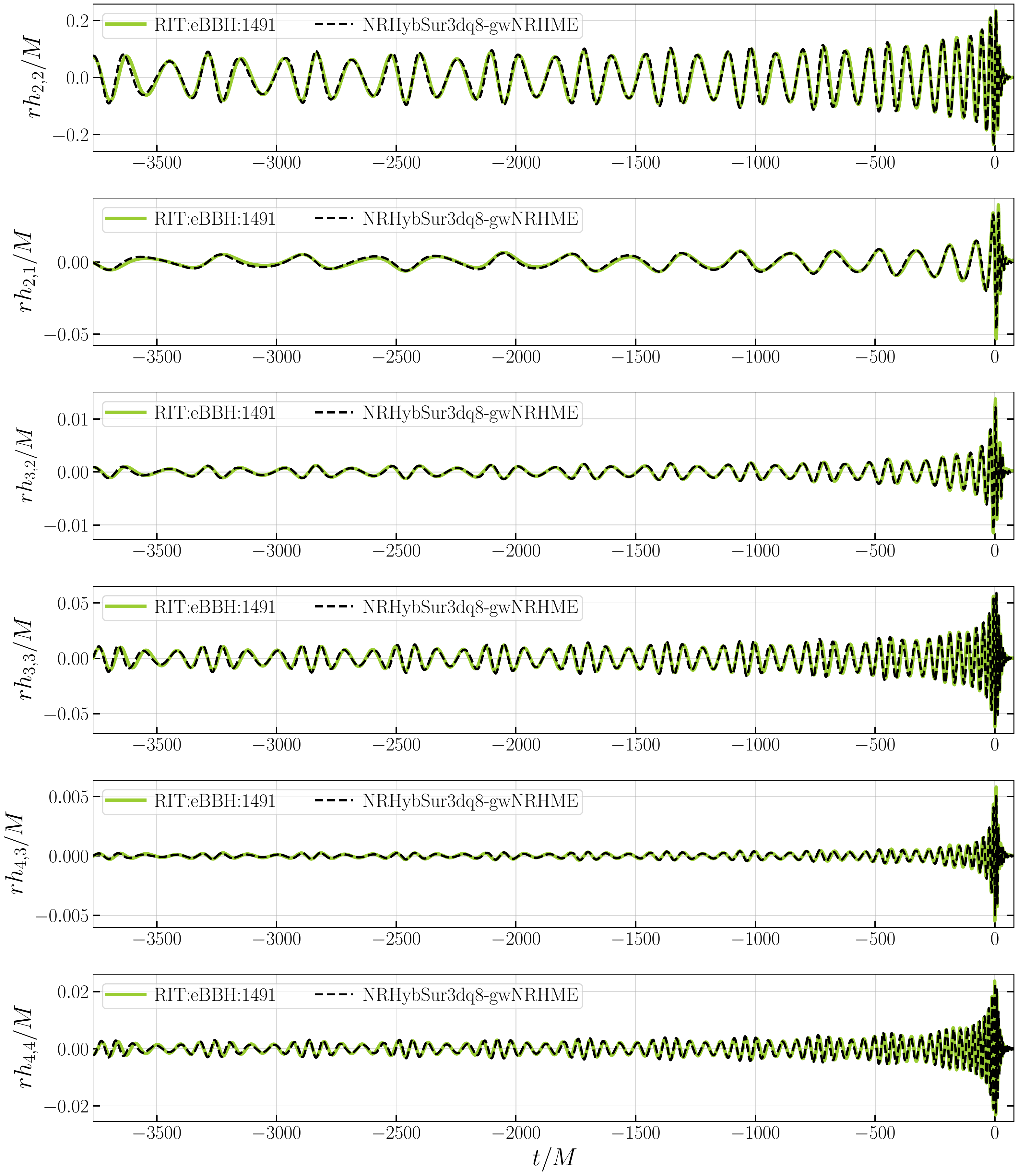}
\caption{We show the eccentric spherical harmonic modes (black dashed lines) obtained 
from the \texttt{NRHybSur3dq8-gwNRHME} model and corresponding NR data from \texttt{RIT:eBBH:1491} simulation (green solid lines). We obtain \texttt{NRHybSur3dq8-gwNRHME} predictions by combining quadrupolar eccentric model \texttt{EccentricIMR} and circular waveform model \texttt{NRHybSur3dq8} using \texttt{gwNRHME} framework (available at \href{https://github.com/tousifislam/gwModels}{https://github.com/tousifislam/gwModels}). This simulation is characterized by mass ratio $q=4$ and eccentricity $e_{\rm ref}=0.19$ measured at the start of the waveform. We find that \texttt{NRHybSur3dq8-gwNRHME} predictions are visually indistinguishable from NR.}
\label{fig:RIT1491_NRHMEcc_wfs}
\end{figure*}

3. For a given mass ratio $q$, eccentricity $e_{\rm ref}$ and mean anomaly $l_{\rm ref}$, we generate quadrupolar eccentric waveform $h^{\rm Ecc}_{2,2}(t; q, e_{\rm ref})$ (using \texttt{EccentricIMR} or \texttt{EccentricTD}) and its corresponding quasi-circular multi-modal waveforms $h^{\rm Cir}_{\ell,m}(t; q, e_{\rm ref}=0)$ (using \texttt{NRHybSur3dq8}). Their initial time-grids may differ, so we cast them onto a common time-grid. We then align them such that the initial orbital phase starts at zero. Subsequently, we use Eq.~(\ref{eq:freq_mod}) and Eq.~(\ref{eq:amp_mod}) to obtain the eccentric modulation $\xi = \xi_{22}^{A}(t;q,e_{\rm ref})$.

\begin{figure*}
\includegraphics[width=\textwidth]{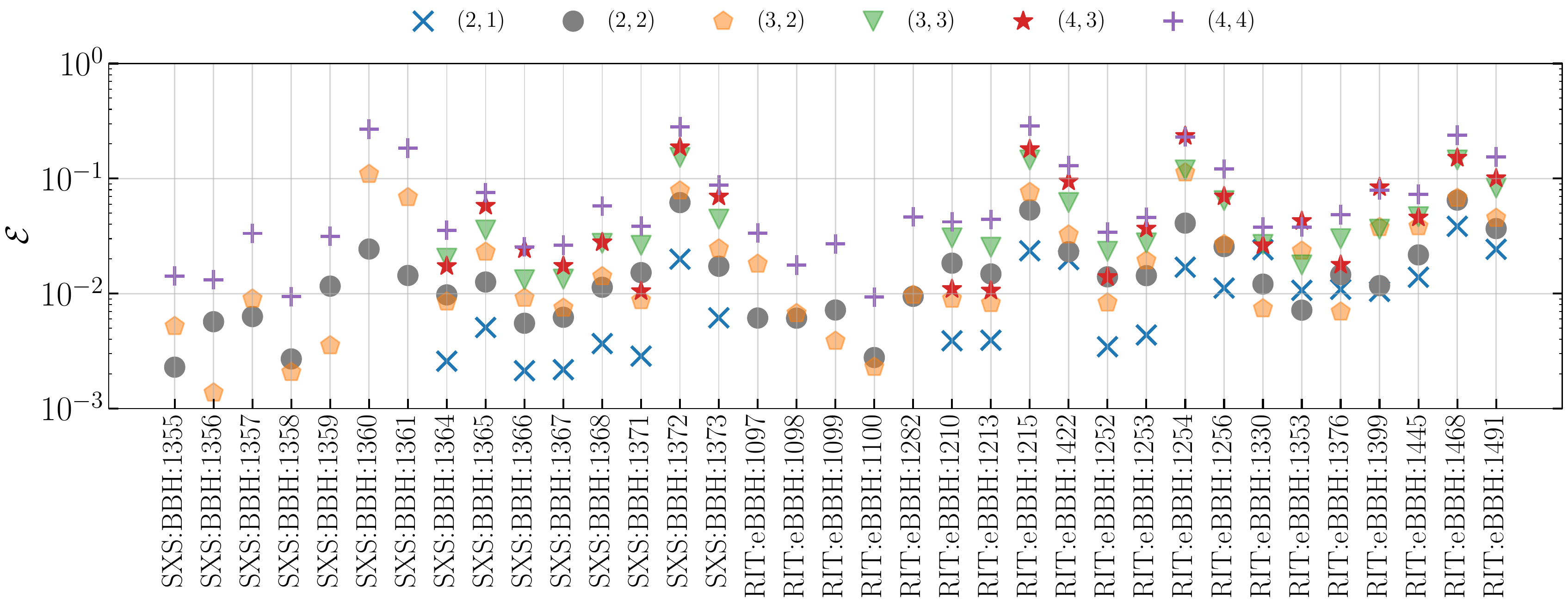}
\caption{We show the relative $L_2$-norm error (defined in Eq.(\ref{eq:l2err})) between the different NR eccentric spherical harmonic mode and corresponding \texttt{EccentricIMR-HM} model prediction. We obtain \texttt{NRHybSur3dq8-gwNRHME} predictions by combining quadrupolar eccentric model \texttt{EccentricIMR} and circular waveform model \texttt{NRHybSur3dq8} using \texttt{gwNRHME} framework (available at \href{https://github.com/tousifislam/gwModels}{https://github.com/tousifislam/gwModels}). For equal-mass binaries, odd $m$ modes become zero due to symmetry and are therefore excluded from this plot. More details are in Section~\ref{sec:l2err}.}
\label{fig:Modelling_errors}
\end{figure*}

4. Finally, using \texttt{gwNRHME} framework, we apply this eccentric modulation on all spherical harmonic modes in $h^{\rm Cir}_{\ell,m}(t; q, e_{\rm ref}=0)$ to convert them into eccentric spherical harmonic modes.

\section{Adding higher order spherical harmonics to eccentric waveform models}
\label{sec:add_HMs}
In this Section, we demonstrate the effectiveness of \texttt{gwNRHME} framework in building efficient non-spinning eccentric multi-modal waveform models. Here, we choose \texttt{EccentricIMR} and \texttt{EccentricTD} models to extract the eccentric modulations in non-spinning binaries. To generate the circular multi-modal waveforms, we choose \texttt{NRHybSur3dq8} (in its non-spinning limit), \texttt{IMRPhenomTHM}~\cite{Estelles:2020twz} (in its non-spinning results) and \texttt{BHPTNRSur1dq1e4} models.

\subsection{Adding higher order spherical harmonics to \texttt{EccentricIMR} model using \texttt{NRHybSur3dq8}}
\label{sec:eccentricimr}
We utilize all 15 SXS NR and 20 RIT NR eccentric simulations considered in Ref.~\cite{Islam:2024tcs}. Initially, we compute the optimized $(2,2)$ mode eccentric waveform $h^{\rm Ecc}_{2,2}(t; q, e_{\rm ref}=0)$ using the \texttt{EccentricIMR} model, with initial eccentricity $e_{\rm ref}$ and mean anomaly $l_{\rm ref}$ values (provided at a reference dimensionless frequency of $x_0=0.07$) that minimize the difference between \texttt{EccentricIMR} model predictions and NR data. These best-fit values of $e_{\rm ref}$ and $l_{\rm ref}$ are given in Ref.~\cite{Islam:2024tcs}. Subsequently, we calculate the initial orbital frequency of the NR data and generate a multi-modal circular non-spinning waveform $h^{\rm Cir}_{\ell,m}(t; q, e_{\rm ref}=0)$ using the \texttt{NRHybSur3dq8} model in its non-spinning limit. We slightly adjust the initial orbital frequencies to ensure that our circular waveform data exceeds the duration of the eccentric NR data. We then utilize the prescription outlined in Section~\ref{sec:gwNRHME} to combine $h^{\rm Cir}_{\ell,m}(t; q, e_{\rm ref}=0)$ and $h^{\rm Ecc}_{2,2}(t; q, e_{\rm ref}, l_{\rm ref})$ through \texttt{gwNRHME}, yielding a multi-modal eccentric waveform. We denote this makeshift model as \texttt{NRHybSur3dq8-gwNRHME}. This model can be seen as an eccentric extension of the \texttt{NRHybSur3dq8} model and as a higher mode extension of the \texttt{EccentricIMR} model.

\begin{figure*}[htb]
	\centering
	\subfigure[]{
		\includegraphics[scale=0.41]{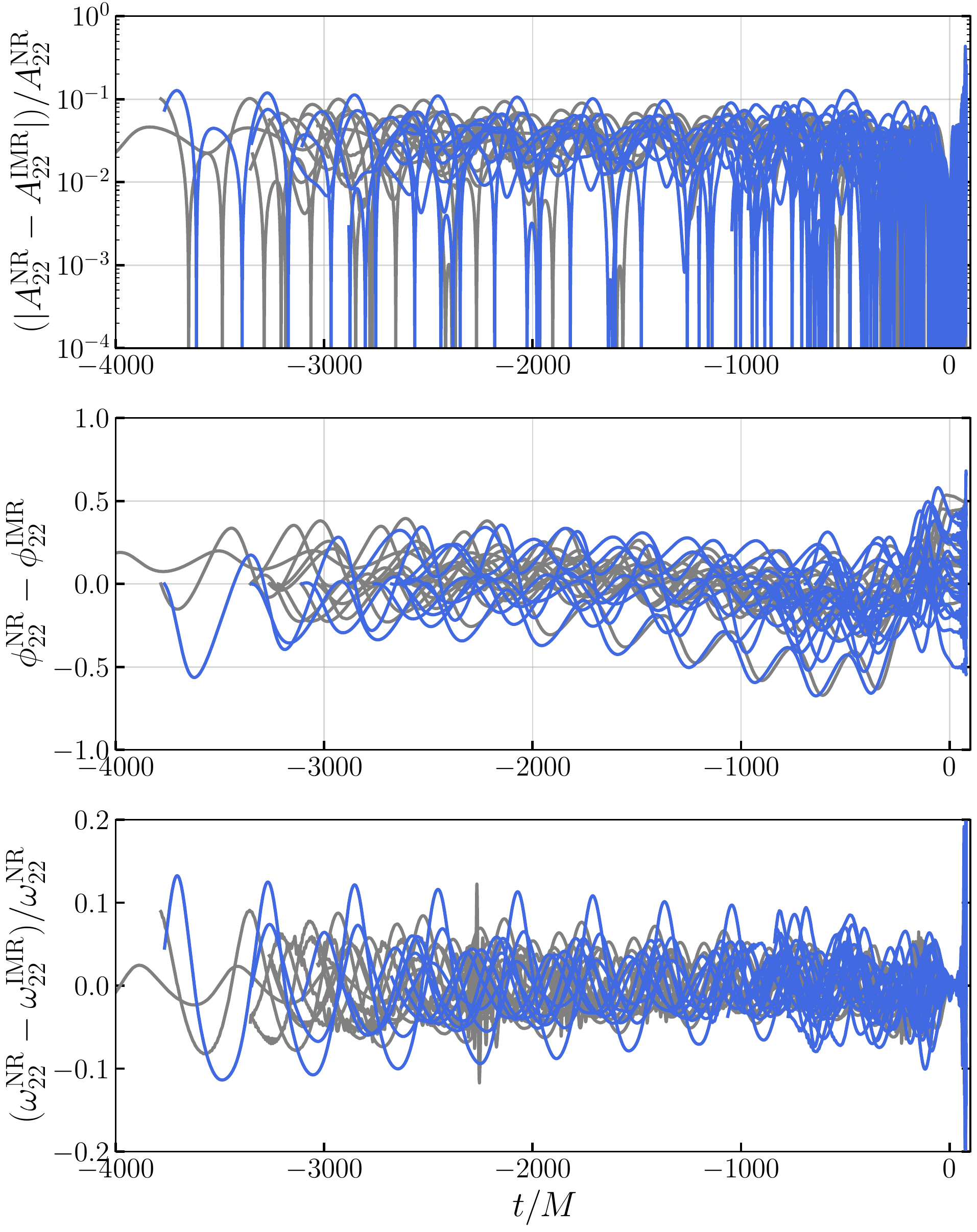}}
	\subfigure[]{
		\includegraphics[scale=0.41]{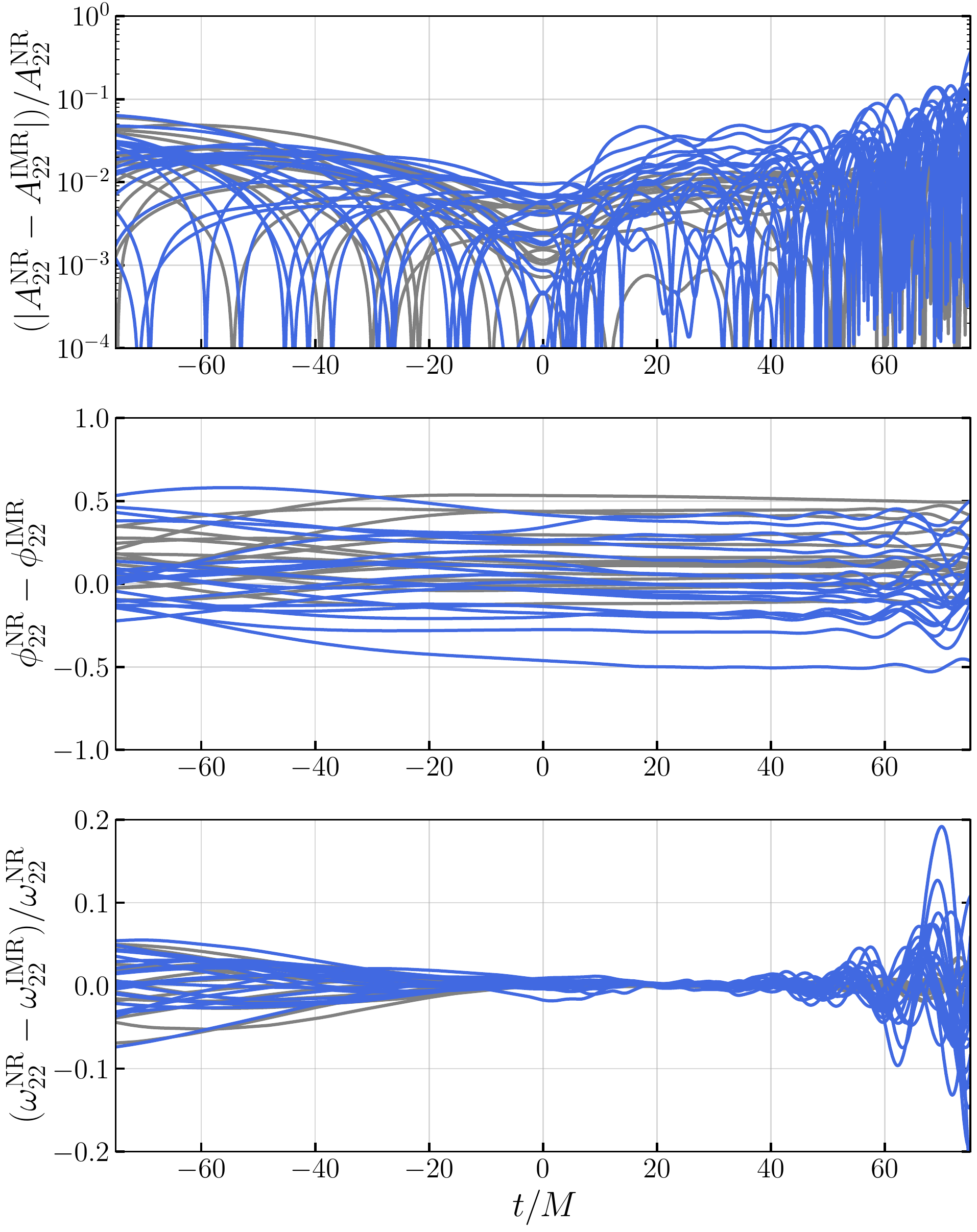}}\\
    \subfigure[]{
		\includegraphics[scale=0.41]{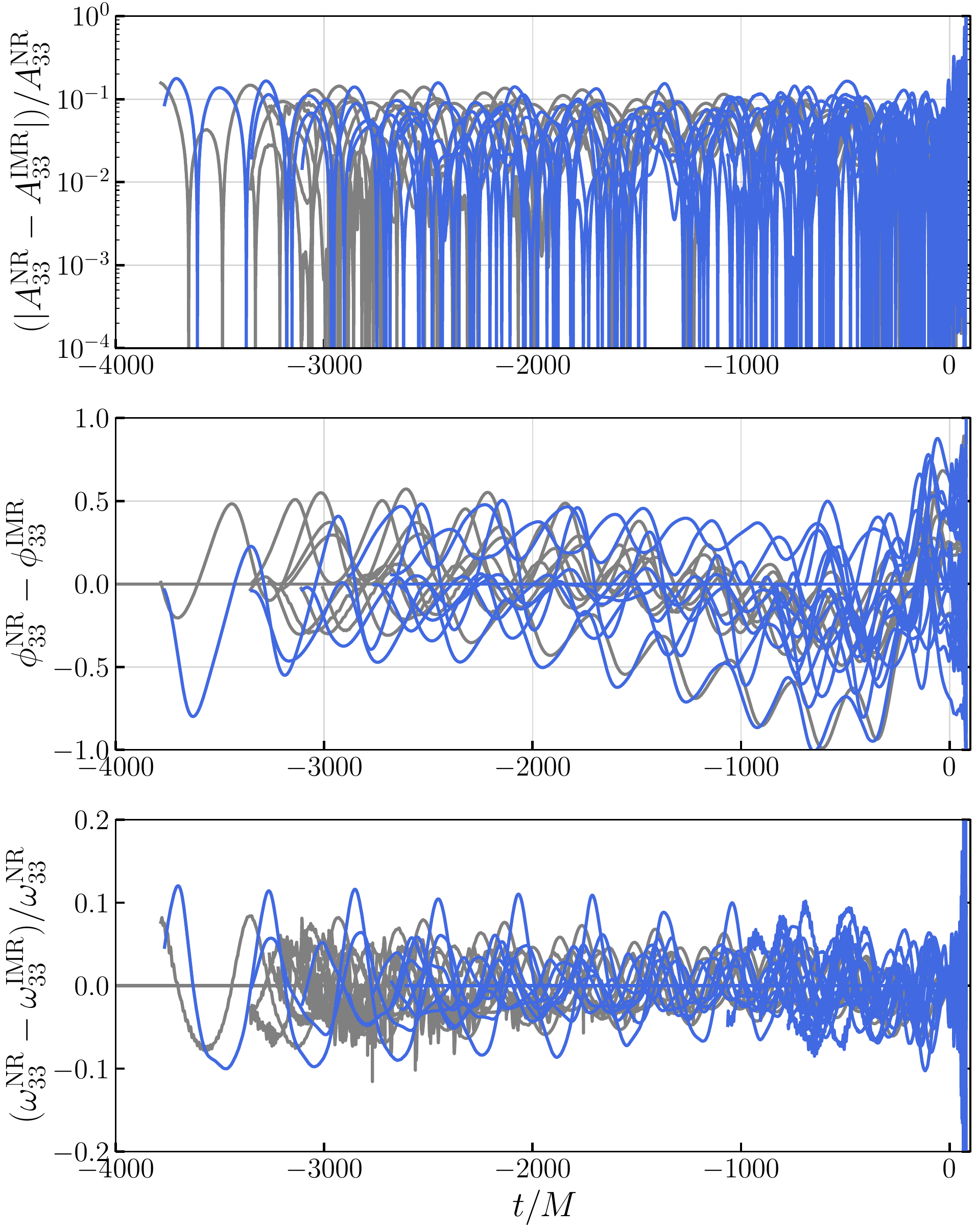}}
	\subfigure[]{
		\includegraphics[scale=0.41]{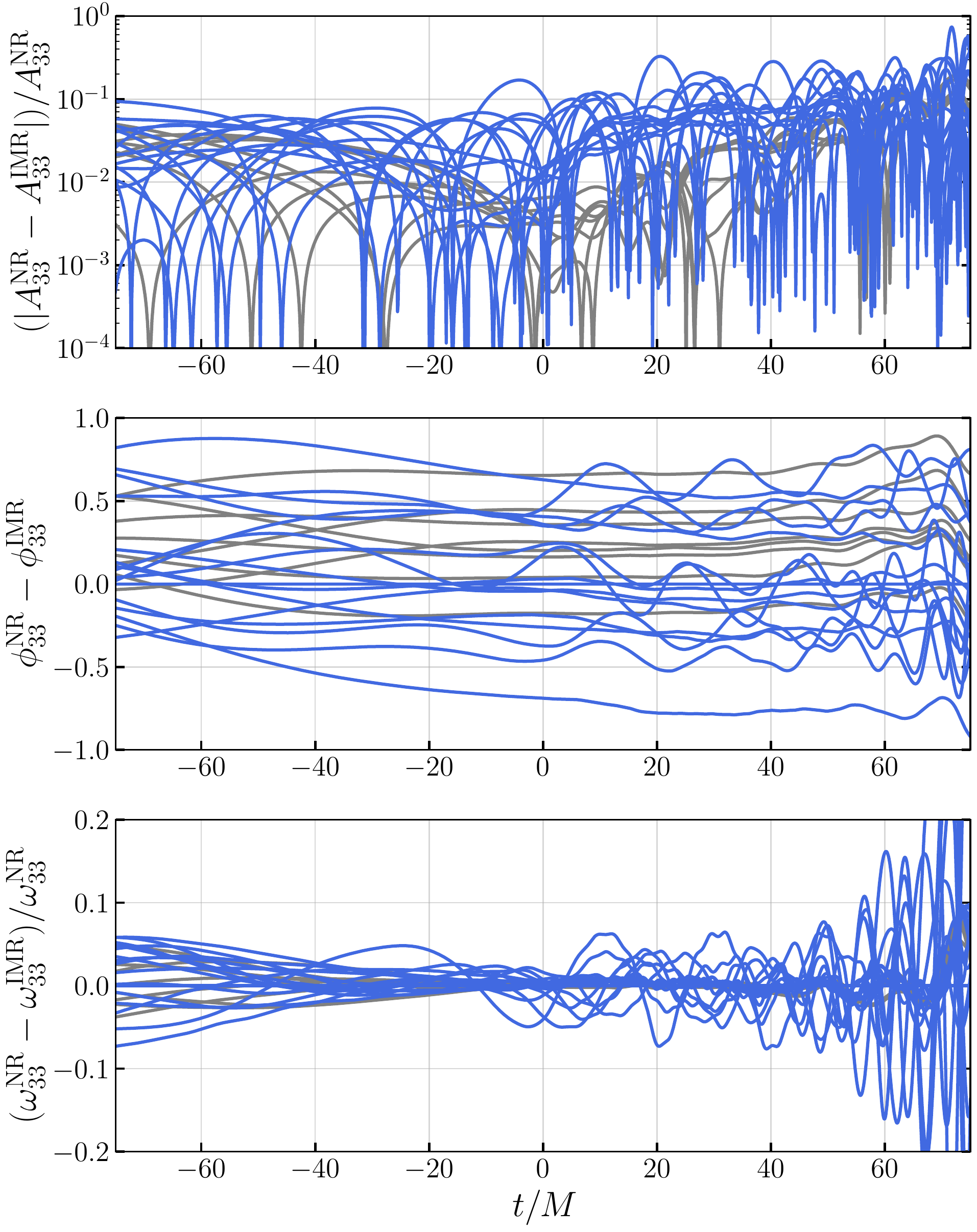}}
	\caption{We show relative amplitude errors (upper panel), absolute phase errors (middle panel) and the relative frequency errors (lower panel) between NR data and \texttt{NRHybSur3dq8-gwNRHME} waveforms for the $(2,2)$ [panel (a) and (b)] and $(3,3)$ [panel (c) and (d)] modes. Blue solid lines denote errors against RIT NR data while grey lines are used for SXS NR data. While panel (a) and (c) focus on the full waveform, panels (b) and (d) only shows the merger-ringdown part. We use red dashed (dotted) lines to indicate median ($95\%$ percentile) error. More details are in Section~\ref{sec:l2err}.}
\label{fig:22_33_mode_errors}
\end{figure*}

\subsubsection{Waveform comparison}
We visually compare \texttt{NRHybSur3dq8-gwNRHME} predictions against all 35 eccentric NR data considered in this study. In each case, we find that combining \texttt{NRHybSur3dq8} and \texttt{EccentricIMR} yields NR-faithful $(2,2)$ spherical harmonic modes. Furthermore,
\texttt{NRHybSur3dq8-gwNRHME} provides accurate higher-order spherical harmonic modes for eccentric BBH mergers. This suggests that combining a NR surrogate waveform model with available PN-based model can help us build waveform models to cover broader parts of the binary parameter space. In many cases where the mass ratio is large ($q \leq 4$) and eccentricity is high ($e_{\rm ref} \leq 0.1$), NR data for higher-order modes exhibit  varying degree of noise, while \texttt{NRHybSur3dq8-gwNRHME} modes are clean and closely follow the best-fit-by-eye curve. As demonstrations, we show mode-by-mode comparison between NR data and respective \texttt{EccentricIMR-HM} waveform in Figures~\ref{fig:SXS1371_NRHMEcc_wfs} and ~\ref{fig:RIT1491_NRHMEcc_wfs} for \texttt{SXS:BBH:1373} and \texttt{RIT:eBBH:1491} respectively. These two simulations represent available high mass ratio and high eccentricity NR simulations in the SXS and RIT catalogs respectively. Their mass ratio values are $q=3$ and $q=4$ while the eccentricities are $0.09$ and $0.19$ measured at $x=0.075$~\cite{Islam:2024tcs}.
We find no visual differences between NR data and \texttt{NRHybSur3dq8-gwNRHME} predictions.

\subsubsection{Time domain errors}
\label{sec:l2err}
We evaluate the modeling accuracy of \texttt{NRHybSur3dq8-gwNRHME} by computing a time/phase optimized time-domain relative $L_2$-norm between the NR data and \texttt{NRHybSur3dq8-gwNRHME} predictions. The relative $L_2$-norm between two waveforms $h_1(t)$ and $h_2(t)$ is given by:
\begin{equation}\label{eq:l2err}
\mathcal{E} = \int_{t_{\rm min}}^{t_{\rm max}} \frac{|h_{1}(t) - h_{2}(t)|^2}{|h_{1}(t)|^2} dt,
\end{equation}
where $t_{\rm min}$ and $t_{\rm max}$ represent the initial and final times of the waveforms. Figure~\ref{fig:Modelling_errors} illustrates the relative $L_2$-norm between the NR data and \texttt{NRHybSur3dq8-gwNRHME} predictions for modes $(2,2)$, $(2,1)$, $(3,2)$, $(3,3)$, $(4,3)$, and $(4,4)$. We observe that the $L_2$-norm error for the \texttt{NRHybSur3dq8-gwNRHME} model remains comparable to that of the \texttt{EccentricIMR} model~\cite{Islam:2024tcs}, particularly as the $(2,2)$ mode remains virtually unaffected by the \texttt{gwNRHME} framework. Relative $L_2$-norm errors for the $(2,2)$ mode is $\sim 10^{-2}$. For the higher order modes, errors slightly increase and reaches $\sim 10^{-1}$ for some cases - specially when the eccentricity increases. Even then errors in $(2,1)$, $(3,3)$ and $(4,4)$ modes are mostly a couple of percent in most binaries considered. We must note that the errors in higher order modes will be dependent on the accuracy of the quadrupolar eccentric waveform model. Given the error in the $(2,2)$ mode is $\sim 10^{-2}$, it makes sense that higher order mode errors are slightly larger. It will therefore be important in future to make sure that the eccentric quadrupolar waveform model is accurate.  Besides, for high eccentricity simulations, NR data also exhibit lot of noise (cf. Figure~6 of Ref.~\cite{Islam:2024rhm}) while \texttt{NRHybSur3dq8-gwNRHME} modes are clean. This may also sometimes wrongly yield a larger $L_2$-norm for the higher order modes.

Next, we calculate the relative amplitude difference
\begin{equation}
\frac{|A_{\ell,m}^{\tt NR} - A_{\ell,m}^{\tt IMR}|}{A_{\ell,m}^{\tt NR}},
\end{equation}
absolute phase difference
\begin{equation}
\phi_{\ell,m}^{\tt NR} - \phi_{\ell,m}^{\tt IMR},
\end{equation}
and relative $(2,2)$ mode frequency difference
\begin{equation}
\frac{\omega_{\ell,m}^{\tt NR} - \omega_{\ell,m}^{\tt IMR}}{\omega_{\ell,m}^{\tt NR}},
\end{equation}
between NR data and the corresponding \texttt{NRHybSur3dq8-gwNRHME} waveforms for all cases. 
\begin{figure}[htb]
\includegraphics[width=\columnwidth]{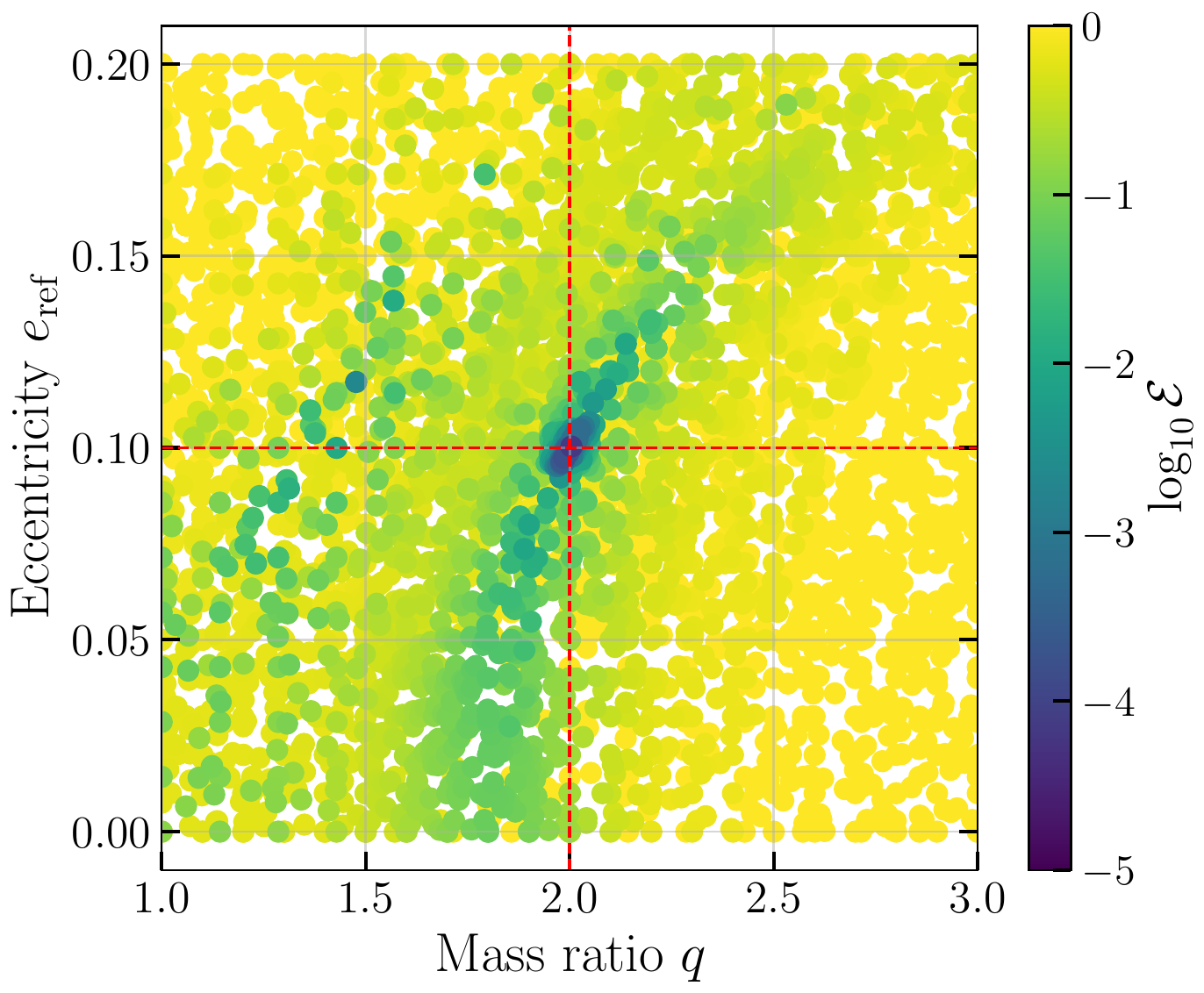}
\caption{We show the relative $L_2$-norm error between reference \texttt{NRHybSur3dq8-gwNRHME} waveform with $[q,e_{\rm ref},l_{\rm ref}]=[2,0.1,0.0]$ and waveforms generated at different mass ratio, eccentricity, mean anomaly values around the parameters associated with the reference waveform. \texttt{NRHybSur3dq8-gwNRHME} waveforms are obtained by combining quadrupolar eccentric model \texttt{EccentricIMR} and circular waveform model \texttt{NRHybSur3dq8} using \texttt{gwNRHME} framework (available at \href{https://github.com/tousifislam/gwModels}{https://github.com/tousifislam/gwModels}). Red dashed line indicate reference waveform point. More details are in Section~\ref{sec:Continuity}.}
\label{fig:l2err_manifold}
\end{figure}
\begin{figure}[htb]
\includegraphics[width=\columnwidth]{mismatches.pdf}
\caption{We show the average frequency-domain mismatches between SXS (blue lines)/RIT (orange lines) NR data and corresponding reference \texttt{NRHybSur3dq8-gwNRHME} waveform predictions obtained using \texttt{gwNRHME} framework (available at \href{https://github.com/tousifislam/gwModels}{https://github.com/tousifislam/gwModels}). Black (red) dashed (dotted) line indicates $1\%$ ($3\%$) mismatch threshold. More details are in Section~\ref{sec:mismatch}.}
\label{fig:mismatch}
\end{figure}
\begin{figure*}
\includegraphics[width=\textwidth]{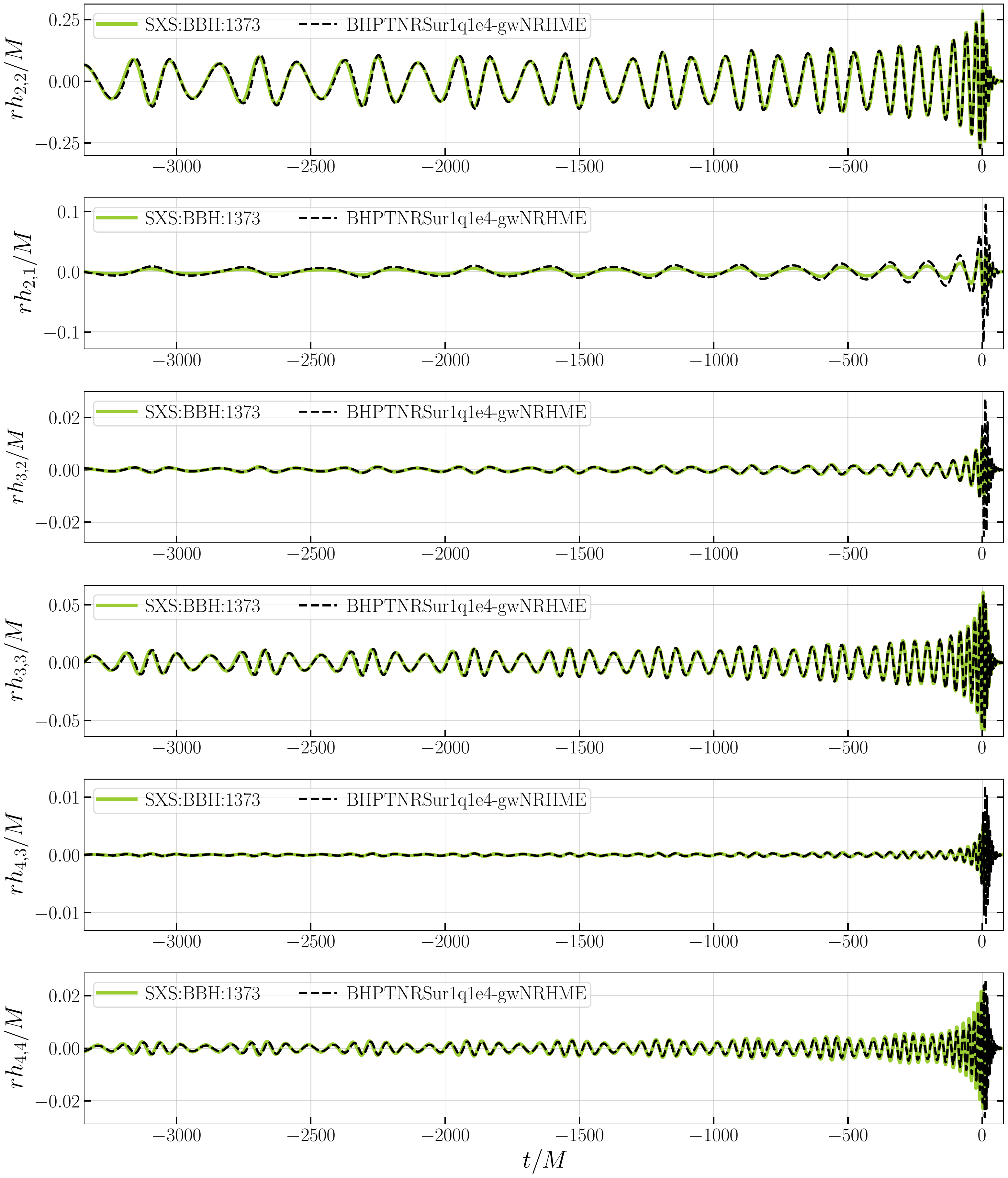}
\caption{We show the eccentric spherical harmonic modes (black dashed lines) obtained 
from the \texttt{BHPTNRSur1dq1e4-gwNRHME} model and corresponding NR data from \texttt{SXS:BBH:1373} simulation (green solid lines). We obtain \texttt{BHPTNRSur1dq1e4-gwNRHME} predictions by combining quadrupolar eccentric model \texttt{EccentricIMR} and circular waveform model \texttt{BHPTNRSur1dq1e4} using \texttt{gwNRHME} framework (available at \href{https://github.com/tousifislam/gwModels}{https://github.com/tousifislam/gwModels}). This simulation is characterized by mass ratio $q=3$ and eccentricity $e_{\rm ref}=0.09$ measured at a reference dimensionless frequency $x_{\rm ref}=0.075$. We find that \texttt{gwNRHME} predictions agrees well with NR.}
\label{fig:SXS1371_NRHMEcc_wfs_bhpt}
\end{figure*}
Here, $A_{\ell,m}^{\rm NR}$ ($A_{\ell,m}^{\tt IMR}$), $\phi_{\ell,m}^{\rm NR}$ ($\phi_{\ell,m}^{\tt IMR}$), and $\omega_{\ell,m}^{\rm NR}$ ($\omega_{\ell,m}^{\tt IMR}$) represent the amplitude, phase, and instantaneous frequency of the NR data (\texttt{NRHybSur3dq8-gwNRHME} waveforms), respectively. In Figure~\ref{fig:22_33_mode_errors}, we show these differences obtained for both SXS and RIT NR data and their corresponding \texttt{NRHybSur3dq8-gwNRHME} counterparts for two representative modes: $(2,2)$ and $(3,3)$. We also illustrate the differences in the full waveform (left panels) and only in the merger-ringdown part (right panels) separately. We find that the phase errors for $(2,2)$ and $(3,3)$ modes remain sub-radian throughout the binary evolution for all modes. The errors in the amplitude is only a couple of percents. However, these errors increase rapidly after $t=50M$. 
For other modes, the qualitative picture remains the same. However, amplitude errors can reach as large as $10\%$ while phase errors are always sub-radian.

\subsubsection{Continuity in parameter space}
\label{sec:Continuity}
To investigate continuity in the behavior of the resultant waveform model, we perform a series of experiments where we compute the $L_2$-norm error between a reference waveform generated with the \texttt{NRHybSur3dq8-gwNRHME} model and waveforms generated at random points in the parameter space around the reference point. To compute the $L_2$-norm error, we consider a total of six modes: $(2,1)$, $(2,2)$, $(3,2)$, $(3,3)$, $(4,3)$, and $(4,4)$. This allows us to probe whether the $L_2$-norm error changes smoothly as we move around in the parameter space. We find that, in al cases, minimum error occurs at the reference point. Otherwise, the errors increase gradually as we move away from the reference point. Furthermore, the behavior of the $L_2$-norm error is similar to the behavior observed for the base \texttt{EccentricIMR} model (shown in Ref.\cite{Islam:2024tcs}). As a demonstration, we show the errors between a reference \texttt{NRHybSur3dq8-gwNRHME} waveform with $(q,e_{\rm ref},l_{\rm ref})=(2,0.1,0.0)$ and waveforms generated at different mass ratio, eccentricity, and mean anomaly values around the parameters associated with the reference waveform in Fig.\ref{fig:l2err_manifold}. 

\subsubsection{Frequency domain mismatches}
\label{sec:mismatch}
Finally, we calculate the frequency-domain mismatches $\mathcal{M}$ between \texttt{NRHybSur3dq8-gwNRHME} and NR data (used in Figure~\ref{fig:Modelling_errors}). The mismatch $\mathcal{M}$ between two waveforms $h_1$ and $h_2$ is defined as~\cite{Cutler:1994ys}),
\begin{equation}
\mathcal{M} =  1 - \left<h_1, h_2\right>,
\end{equation}
with
\begin{gather}
	\left<h_1, h_2\right> = 4 \mathrm{Re}
	\int_{f_{\mathrm{min}}}^{f_{\mathrm{max}}}
	\frac{\tilde{h}_1 (f) \tilde{h}_2^* (f) }{S_n (f)} df,
	\label{Eq:freq_domain_Mismatch}
\end{gather}
where $\tilde{h}(f)$ indicates the Fourier transform of the strain $h(t)$, $^*$ indicates complex conjugation, $\mathrm{Re}$ indicates the real part, and $S_n(f)$ is the one-sided power spectral density (PSD). We adopt the design sensitivity power spectral density (PSD) of the Advanced LIGO detector~\cite{KAGRA:2013rdx}, using a frequency range from $f_{\mathrm{min}} = 20$ Hz to $f_{\mathrm{max}} = 999$ Hz. For mismatch computations, we include a total of 12 spherical harmonic modes: $(\ell, m) = [(2, \pm1), (2, \pm2), (3, \pm2), (3, \pm3), (4, \pm3), (4, \pm4)]$. Additionally, mismatches are evaluated across a range of total binary masses, $M = [40, 80, 120, 160, 200] M_{\odot}$, and we report the average mismatch computed over five different inclination angles, $\iota = [0, \pi/6, \pi/4, \pi/3, \pi/2]$ (Fig.~\ref{fig:mismatch}).

We find that the mismatches between our model and both SXS and RIT NR simulations are typically below $0.01$, and always remain $\leq 0.03$. Notably, the mismatch behavior is consistent across both the SXS and RIT NR datasets. We emphasize that the eccentric modulations in our waveform model are derived from the \texttt{EccentricIMR} waveform, which incorporates conservative dynamics up to 3PN order and radiation-reaction effects up to 2PN order~\cite{Hinder:2008kv}. In the future, mismatches could be further reduced by incorporating higher-order PN corrections to the eccentric quadrupolar mode, or by including eccentric modulations from TEOB/SEOB models or directly from NR data. We leave these improvements for future work.

\subsection{Adding higher order spherical harmonics to \texttt{EccentricIMR} model using \texttt{BHPTNRSur1dq1e4}}
\label{sec:eccentricimr-bhpt}
To demonstrate the modularity of our framework and its compatibility with various existing multi-modal quasi-circular models, we construct another model variant named \texttt{BHPTNRSur1dq1e4-gwNRHME}. In this model, we substitute the \texttt{NRHybSur3dq8} model with \texttt{BHPTNRSur1dq1e4}~\cite{Islam:2022laz}, a surrogate model based on black-hole perturbation theory and NR. The \texttt{BHPTNRSur1dq1e4} model covers mass ratios ranging from $q=2.5$ to $q=10000$ and includes modes up to $\ell=10$. It has been observed that \texttt{BHPTNRSur1dq1e4} provides accurate waveforms up to $q\leq4$, beyond which the model errors become $0.001$ or less. This implies that for $q\leq4$, where \texttt{BHPTNRSur1dq1e4} errors are larger, especially in the higher-order modes, the resulting accuracy of the \texttt{BHPTNRSur1dq1e4-gwNRHME} model will also be compromised. Nevertheless, our analysis reveals that the predictions of \texttt{BHPTNRSur1dq1e4-gwNRHME} align well with NR simulations for $q\ge3$ for the $(2,2)$, $(3,3)$ and $(4,4)$ modes. For other modes, the errors are expected to decrease as the mass ratio increases. However we do not have enough eccentric NR data for $q\geq4$. 

In Figure~\ref{fig:SXS1371_NRHMEcc_wfs_bhpt}, we present the eccentric spherical harmonic modes obtained from the \texttt{EccentricIMR-HM} model alongside the corresponding NR data from the \texttt{SXS:BBH:1373} simulations. We observe that, with the exception of the $(\ell,m)=(2,1)$ mode, the predictions of \texttt{EccentricIMR-HM} are visually agrees well with the NR data. This serves as a demonstration that in order to deliver NR-faithful eccentric waveform models through \texttt{gwNRHME}, it is crucial to have both accurate multi-modal circular waveform models and accurate eccentric quadrupolar waveform models. Inaccuracies in either of these two models will propagate to the resultant multi-modal eccentric waveform model.

\subsection{Adding higher order spherical harmonics to \texttt{EccentricIMR} model using \texttt{IMRPhenomTHM}}
\label{sec:eccentricimr-phenom}
Our next model is constructed by combining the eccentric component of the quadrupolar waveform model \texttt{EccentricIMR} with the multi-modal phenomenological time-domain circular waveform model named \texttt{IMRPhenomTHM}~\cite{Estelles:2020twz}. The latter includes $(2,2)$, $(2,1)$, $(3,3)$, $(4,4)$, and $(5,5)$ modes. We denote the combined model as \texttt{IMRPhenomTHM-gwNRHME}. We observe that \texttt{IMRPhenomTHM-gwNRHME} exhibits similar accuracies and behaviors as observed in the \texttt{NRHybSur3dq8-gwNRHME} and \texttt{BHPTNRSur1dq1e4-gwNRHME} models, except for the $(2,1)$ mode, where the accuracy of the \texttt{IMRPhenomTHM-gwNRHME} model drops significantly. As the results are otherwise similar, we do not show the resultant waveforms here. However, \texttt{IMRPhenomTHM-gwNRHME} model is accessible through the \texttt{gwModels} package.

\subsection{Adding higher order spherical harmonics to \texttt{EccentricTD} model using \texttt{NRHybSr3dq8}}
\label{sec:eccentrictd}
While our results with the \texttt{EccentricIMR} model clearly demonstrate the power of using observed universal relations in eccentric waveforms~\cite{Islam:2024tcs} and the associated \texttt{gwNRHME} framework~\cite{Islam:2024tcs} to efficiently convert multi-modal circular waveform models into multi-modal eccentric waveform models if the quadrupolar eccentric model is known, for completeness, we also consider the \texttt{EccentricTD} model. This model is an inspiral-only waveform and therefore comes with added complexity. Furthermore, it only includes PN corrections in the phase and not in the amplitude, making it less accurate than \texttt{EccentricIMR} model. Nonetheless, we demonstrate that \texttt{gwNRHME} can still provide higher-order eccentric spherical harmonic modes for the \texttt{EccentricTD} model using the same strategies outlined in Section~\ref{sec:gwNRHME}.

\begin{figure*}
\includegraphics[width=\textwidth]{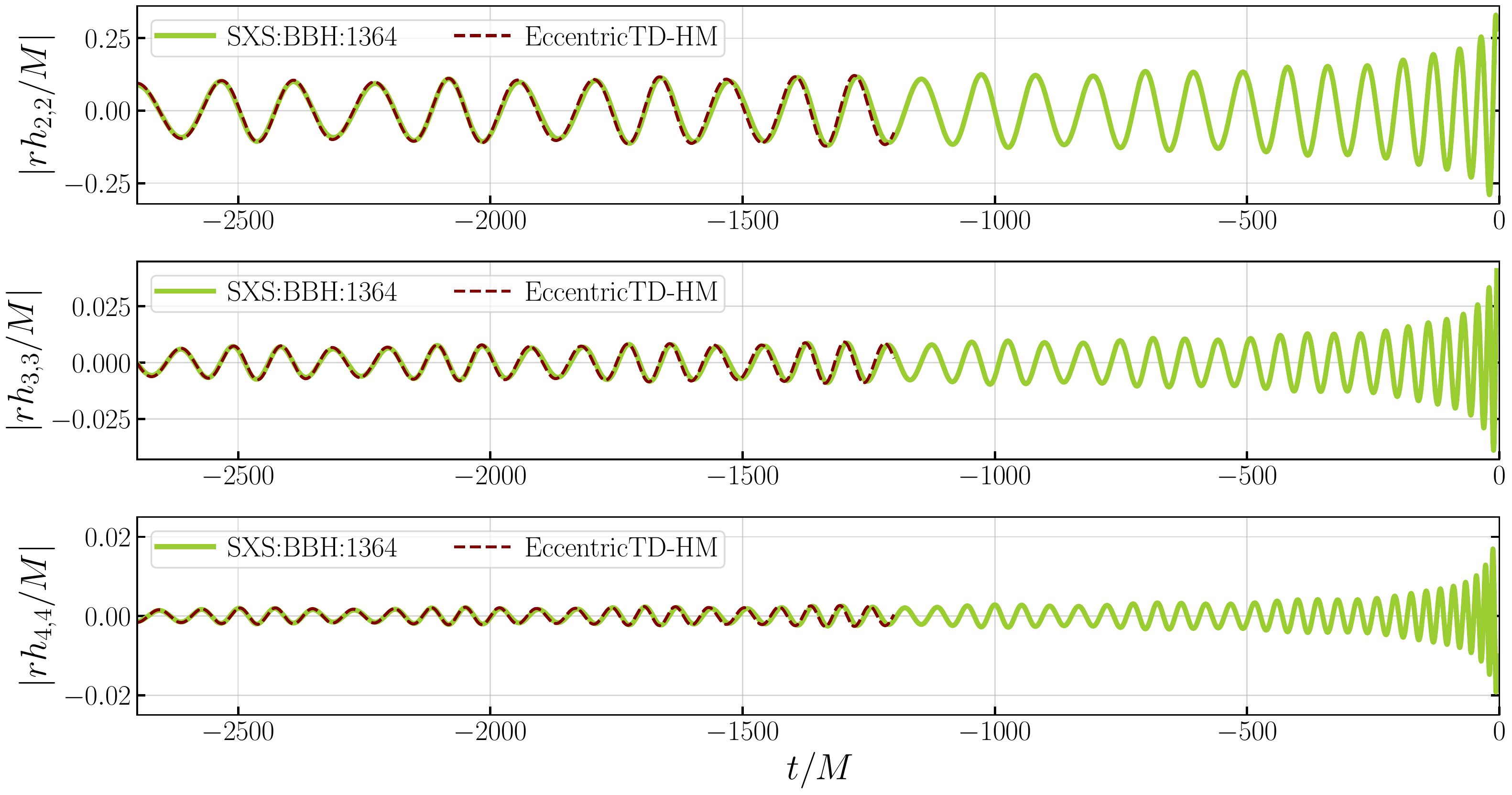}
\caption{We show the eccentric spherical harmonic modes (black dashed lines) obtained 
from the \texttt{EccentricTD-HM} model and corresponding NR data from \texttt{SXS:BBH:1364} simulation (green solid lines). We obtain \texttt{EccentricTD-HM} predictions by combining quadrupolar eccentric model \texttt{EccentricTD} and circular waveform model \texttt{NRHybSur3dq8} using \texttt{gwNRHME} framework (available at \href{https://github.com/tousifislam/gwModels}{https://github.com/tousifislam/gwModels}). This simulation is characterized by mass ratio $q=2$ and eccentricity $e_{\rm ref}=0.05$ measured at a reference dimensionless frequency $x_{\rm ref}=0.075$. We find that \texttt{gwNRHME} predictions are visually almost indistinguishable from NR in the inspiral.}
\label{fig:EccentricTDHM}
\end{figure*}

Figure~\ref{fig:EccentricTDHM} shows the eccentric spherical harmonic modes obtained 
from the \texttt{EccentricTD-HM} model and corresponding NR data from \texttt{SXS:BBH:1364} simulation. We obtain \texttt{EccentricTD-HM} predictions by combining quadrupolar eccentric model \texttt{EccentricTD} and circular waveform model \texttt{NRHybSur3dq8} using \texttt{gwNRHME} framework. This simulation is characterized by mass ratio $q=2$ and eccentricity $e_{\rm ref}=0.05$ measured at a reference dimensionless frequency $x_{\rm ref}=0.075$. We find that \texttt{gwNRHME} predictions are visually almost indistinguishable from NR in the inspiral. We do not show \texttt{EccentricTD-HM} waveforms after $t=-1200M$ as the base PN \texttt{EccentricTD} model predictions shows significant dephasing. We fnd that the relative $L_2$-norm error between NR data and \texttt{EccentricTD-HM} model is 0.01, 0.03 and 0.07 for the $(2,2)$, $(3,3)$ and $(4,4)$ mode respectively if we consider data up to $t=-1200M$.

\section{Discussion and conclusion}
\label{sec:conclusion}
In Ref.~\cite{Islam:2024rhm}, using publicly available eccentric NR simulations, we have demonstrated that the phenomenology of gravitational waveform in eccentric BBH mergers are significantly simpler. Different quantities of interest (for example amplitudes, phases and frequencies) in different spherical harmonic modes can be modelled by adding a single eccentric modulation on top of the quasi-circular expectations. This reduces the modelling choices drastically resulting an easier way to develop mulit-modal eccentric BBH waveform model. Our framework is modular and can in principle be used to extend any quadrupolar eccentric waveform model immediately. This framework is named as \texttt{gwNRHME} and and can be accessed at \href{https://github.com/tousifislam/gwModels}{https://github.com/tousifislam/gwModels}.

In this paper, we showcase the effectiveness of \texttt{gwNRHME} in transforming quasi-circular multi-modal waveform models into eccentric multi-modal waveform models using only a quadrupolar eccentric model. Specifically, we select the publicly available eccentric inspiral-merger-ringdown model named \texttt{EccentricIMR}~\cite{Hinder:2017sxy} and convert it into an eccentric multi-modal model by combining it with the quasi-circular multi-modal model \texttt{NRHybSur3dq8}, \texttt{BHPTNRSur1dq1e4} and \texttt{IMRPhenomTHM} through \texttt{gwNRHME} framework. We refer to the resulting models as \texttt{NRHybSur3dq8-gwNRHME} \texttt{BHPTNRSur1dq1e4-gwNRHME} and \texttt{IMRPhenomTHM-gwNRHME}. We then demonstrate their reasonable accuracy in matching NR data. Subsequently, we employ the same approach with \texttt{EccentricTD}~\cite{Tanay:2016zog}, one of the most popular inspiral-only eccentric non-spinning waveform models, and illustrate its extension to higher-order modes.

Our method can be employed to assess the accuracy and, in some cases, validate the higher-order modes in existing multi-modal eccentric waveform models, particularly those within the effective-one-body frameworks, such as \texttt{SEOBNRv4HME}, \texttt{SEOBNRE} and \texttt{TEOBReSumS}. We leave this exploration for the near future.

We must note that since \texttt{gwNRHME} combines two different waveform models—one multi-modal circular waveform model and one quadrupolar eccentric waveform model—the accuracy of the resulting multi-modal eccentric waveform model will depend on the accuracy of the constituent models themselves. Therefore, it is crucial to develop accurate multi-modal circular models as well as accurate quadrupolar eccentric models in the future to ensure the seamless generation of multi-modal eccentric waveform models. This may entail incorporating higher-order eccentric PN corrections during the construction of quadrupolar eccentric models or conducting additional eccentric NR simulations to facilitate the development of data-driven quadrupolar eccentric models.

One of the current limitations of of our framework is the availability of fast qudrupolar eccentric waveform model. A possible remedy is to build a reduced order approximation or fast phenomenological model for the eccentric modulations based on either PN, NR, EOB or a combination of these in future.

\begin{acknowledgments}
We thank Chandra Kant Mishra, Frank Ohme, Ajit Kumar Mehta, Harald P. Pfeiffer and Tejaswi Venumadhav for fruitful discussions. We are also thankful to Lorenzo Pompili for carefully reading our manuscript and for his suggestions. We are grateful to the SXS collaboration and RIT NR group for maintaining publicly available catalog of NR simulations which has been used in this study.
This research was supported in part by the National Science Foundation under Grant No. NSF PHY-2309135 and the Simons Foundation (216179, LB). Most of this work was conducted on the UMass-URI UNITY supercomputer supported by the Massachusetts Green High-Performance Computing Center (MGHPCC) and CARNiE at the Center for Scientific Computing and Data Science Research (CSCDR) of UMassD, which is supported by  ONR/DURIP grant no.\ N00014181255.
\end{acknowledgments}

\bibliography{References}

@article{Hinder:2017sxy,
    author = "Hinder, Ian and Kidder, Lawrence E. and Pfeiffer, Harald P.",
    title = "{Eccentric binary black hole inspiral-merger-ringdown gravitational waveform model from numerical relativity and post-Newtonian theory}",
    eprint = "1709.02007",
    archivePrefix = "arXiv",
    primaryClass = "gr-qc",
    doi = "10.1103/PhysRevD.98.044015",
    journal = "Phys. Rev. D",
    volume = "98",
    number = "4",
    pages = "044015",
    year = "2018"
}

@article{Memmesheimer:2004cv,
    author = "Memmesheimer, Raoul-Martin and Gopakumar, Achamveedu and Schaefer, Gerhard",
    title = "{Third post-Newtonian accurate generalized quasi-Keplerian parametrization for compact binaries in eccentric orbits}",
    eprint = "gr-qc/0407049",
    archivePrefix = "arXiv",
    doi = "10.1103/PhysRevD.70.104011",
    journal = "Phys. Rev. D",
    volume = "70",
    pages = "104011",
    year = "2004"
}

@article{Konigsdorffer:2006zt,
    author = "Konigsdorffer, Christian and Gopakumar, Achamveedu",
    title = "{Phasing of gravitational waves from inspiralling eccentric binaries at the third-and-a-half post-Newtonian order}",
    eprint = "gr-qc/0603056",
    archivePrefix = "arXiv",
    doi = "10.1103/PhysRevD.73.124012",
    journal = "Phys. Rev. D",
    volume = "73",
    pages = "124012",
    year = "2006"
}

@article{Damour:2004bz,
    author = "Damour, Thibault and Gopakumar, Achamveedu and Iyer, Bala R.",
    title = "{Phasing of gravitational waves from inspiralling eccentric binaries}",
    eprint = "gr-qc/0404128",
    archivePrefix = "arXiv",
    doi = "10.1103/PhysRevD.70.064028",
    journal = "Phys. Rev. D",
    volume = "70",
    pages = "064028",
    year = "2004"
}

@article{Tiwari:2019jtz,
    author = "Tiwari, Srishti and Achamveedu, Gopakumar and Haney, Maria and Hemantakumar, Phurailatapam",
    title = "{Ready-to-use Fourier domain templates for compact binaries inspiraling along moderately eccentric orbits}",
    eprint = "1905.07956",
    archivePrefix = "arXiv",
    primaryClass = "gr-qc",
    doi = "10.1103/PhysRevD.99.124008",
    journal = "Phys. Rev. D",
    volume = "99",
    number = "12",
    pages = "124008",
    year = "2019"
}

@article{Huerta:2014eca,
    author = "Huerta, E. A. and Kumar, Prayush and McWilliams, Sean T. and O'Shaughnessy, Richard and Yunes, Nicol\'as",
    title = "{Accurate and efficient waveforms for compact binaries on eccentric orbits}",
    eprint = "1408.3406",
    archivePrefix = "arXiv",
    primaryClass = "gr-qc",
    doi = "10.1103/PhysRevD.90.084016",
    journal = "Phys. Rev. D",
    volume = "90",
    number = "8",
    pages = "084016",
    year = "2014"
}

@article{Moore:2016qxz,
    author = "Moore, Blake and Favata, Marc and Arun, K. G. and Mishra, Chandra Kant",
    title = "{Gravitational-wave phasing for low-eccentricity inspiralling compact binaries to 3PN order}",
    eprint = "1605.00304",
    archivePrefix = "arXiv",
    primaryClass = "gr-qc",
    reportNumber = "LIGO-DCC-P1500268",
    doi = "10.1103/PhysRevD.93.124061",
    journal = "Phys. Rev. D",
    volume = "93",
    number = "12",
    pages = "124061",
    year = "2016"
}

@article{Cho:2021oai,
    author = "Cho, Gihyuk and Tanay, Sashwat and Gopakumar, Achamveedu and Lee, Hyung Mok",
    title = "{Generalized quasi-Keplerian solution for eccentric, nonspinning compact binaries at 4PN order and the associated inspiral-merger-ringdown waveform}",
    eprint = "2110.09608",
    archivePrefix = "arXiv",
    primaryClass = "gr-qc",
    doi = "10.1103/PhysRevD.105.064010",
    journal = "Phys. Rev. D",
    volume = "105",
    number = "6",
    pages = "064010",
    year = "2022"
}

@article{Chattaraj:2022tay,
    author = "Chattaraj, Abhishek and RoyChowdhury, Tamal and Divyajyoti and Mishra, Chandra Kant and Gupta, Anshu",
    title = "{High accuracy post-Newtonian and numerical relativity comparisons involving higher modes for eccentric binary black holes and a dominant mode eccentric inspiral-merger-ringdown model}",
    eprint = "2204.02377",
    archivePrefix = "arXiv",
    primaryClass = "gr-qc",
    reportNumber = "LIGO-P2200106",
    doi = "10.1103/PhysRevD.106.124008",
    journal = "Phys. Rev. D",
    volume = "106",
    number = "12",
    pages = "124008",
    year = "2022"
}

@article{Ramos-Buades:2021adz,
    author = "Ramos-Buades, Antoni and Buonanno, Alessandra and Khalil, Mohammed and Ossokine, Serguei",
    title = "{Effective-one-body multipolar waveforms for eccentric binary black holes with nonprecessing spins}",
    eprint = "2112.06952",
    archivePrefix = "arXiv",
    primaryClass = "gr-qc",
    doi = "10.1103/PhysRevD.105.044035",
    journal = "Phys. Rev. D",
    volume = "105",
    number = "4",
    pages = "044035",
    year = "2022"
}

@article{Chiaramello:2020ehz,
    author = "Chiaramello, Danilo and Nagar, Alessandro",
    title = "{Faithful analytical effective-one-body waveform model for spin-aligned, moderately eccentric, coalescing black hole binaries}",
    eprint = "2001.11736",
    archivePrefix = "arXiv",
    primaryClass = "gr-qc",
    doi = "10.1103/PhysRevD.101.101501",
    journal = "Phys. Rev. D",
    volume = "101",
    number = "10",
    pages = "101501",
    year = "2020"
}

@article{Riemenschneider:2021ppj,
    author = "Riemenschneider, Gunnar and Rettegno, Piero and Breschi, Matteo and Albertini, Angelica and Gamba, Rossella and Bernuzzi, Sebastiano and Nagar, Alessandro",
    title = "{Assessment of consistent next-to-quasicircular corrections and postadiabatic approximation in effective-one-body multipolar waveforms for binary black hole coalescences}",
    eprint = "2104.07533",
    archivePrefix = "arXiv",
    primaryClass = "gr-qc",
    doi = "10.1103/PhysRevD.104.104045",
    journal = "Phys. Rev. D",
    volume = "104",
    number = "10",
    pages = "104045",
    year = "2021"
}

@article{Albanesi:2022xge,
    author = "Albanesi, Simone and Placidi, Andrea and Nagar, Alessandro and Orselli, Marta and Bernuzzi, Sebastiano",
    title = "{New avenue for accurate analytical waveforms and fluxes for eccentric compact binaries}",
    eprint = "2203.16286",
    archivePrefix = "arXiv",
    primaryClass = "gr-qc",
    doi = "10.1103/PhysRevD.105.L121503",
    journal = "Phys. Rev. D",
    volume = "105",
    number = "12",
    pages = "L121503",
    year = "2022"
}

@article{Albanesi:2023bgi,
    author = "Albanesi, Simone and Bernuzzi, Sebastiano and Damour, Thibault and Nagar, Alessandro and Placidi, Andrea",
    title = "{Faithful effective-one-body waveform of small-mass-ratio coalescing black hole binaries: The eccentric, nonspinning case}",
    eprint = "2305.19336",
    archivePrefix = "arXiv",
    primaryClass = "gr-qc",
    doi = "10.1103/PhysRevD.108.084037",
    journal = "Phys. Rev. D",
    volume = "108",
    number = "8",
    pages = "084037",
    year = "2023"
}

@article{Setyawati:2021gom,
    author = "Setyawati, Yoshinta and Ohme, Frank",
    title = "{Adding eccentricity to quasicircular binary-black-hole waveform models}",
    eprint = "2101.11033",
    archivePrefix = "arXiv",
    primaryClass = "gr-qc",
    doi = "10.1103/PhysRevD.103.124011",
    journal = "Phys. Rev. D",
    volume = "103",
    number = "12",
    pages = "124011",
    year = "2021"
}

@article{Wang:2023ueg,
    author = "Wang, Hao and Zou, Yuan-Chuan and Liu, Yu",
    title = "{Phenomenological relationship between eccentric and quasicircular orbital binary black hole waveform}",
    eprint = "2302.11227",
    archivePrefix = "arXiv",
    primaryClass = "gr-qc",
    doi = "10.1103/PhysRevD.107.124061",
    journal = "Phys. Rev. D",
    volume = "107",
    number = "12",
    pages = "124061",
    year = "2023"
}

@article{Liu:2023ldr,
    author = "Liu, Xiaolin and Cao, Zhoujian and Zhu, Zong-Hong",
    title = "{Effective-One-Body Numerical-Relativity waveform model for Eccentric spin-precessing binary black hole coalescence}",
    eprint = "2310.04552",
    archivePrefix = "arXiv",
    primaryClass = "gr-qc",
    month = "10",
    year = "2023"
}

@article{Joshi:2022ocr,
    author = "Joshi, Abhishek V. and Rosofsky, Shawn G. and Haas, Roland and Huerta, E. A.",
    title = "{Numerical relativity higher order gravitational waveforms of eccentric, spinning, nonprecessing binary black hole mergers}",
    eprint = "2210.01852",
    archivePrefix = "arXiv",
    primaryClass = "gr-qc",
    doi = "10.1103/PhysRevD.107.064038",
    journal = "Phys. Rev. D",
    volume = "107",
    number = "6",
    pages = "064038",
    year = "2023"
}

@article{Huerta:2017kez,
    author = "Huerta, E. A. and others",
    title = "{Eccentric, nonspinning, inspiral, Gaussian-process merger approximant for the detection and characterization of eccentric binary black hole mergers}",
    eprint = "1711.06276",
    archivePrefix = "arXiv",
    primaryClass = "gr-qc",
    doi = "10.1103/PhysRevD.97.024031",
    journal = "Phys. Rev. D",
    volume = "97",
    number = "2",
    pages = "024031",
    year = "2018"
}

@article{Huerta:2016rwp,
    author = "Huerta, E. A. and others",
    title = "{Complete waveform model for compact binaries on eccentric orbits}",
    eprint = "1609.05933",
    archivePrefix = "arXiv",
    primaryClass = "gr-qc",
    doi = "10.1103/PhysRevD.95.024038",
    journal = "Phys. Rev. D",
    volume = "95",
    number = "2",
    pages = "024038",
    year = "2017"
}

@article{Cao:2017ndf,
    author = "Cao, Zhoujian and Han, Wen-Biao",
    title = "{Waveform model for an eccentric binary black hole based on the effective-one-body-numerical-relativity formalism}",
    eprint = "1708.00166",
    archivePrefix = "arXiv",
    primaryClass = "gr-qc",
    doi = "10.1103/PhysRevD.96.044028",
    journal = "Phys. Rev. D",
    volume = "96",
    number = "4",
    pages = "044028",
    year = "2017"
}

@article{Hinderer:2017jcs,
    author = "Hinderer, Tanja and Babak, Stanislav",
    title = "{Foundations of an effective-one-body model for coalescing binaries on eccentric orbits}",
    eprint = "1707.08426",
    archivePrefix = "arXiv",
    primaryClass = "gr-qc",
    doi = "10.1103/PhysRevD.96.104048",
    journal = "Phys. Rev. D",
    volume = "96",
    number = "10",
    pages = "104048",
    year = "2017"
}

@article{Islam:2021mha,
    author = "Islam, Tousif and Varma, Vijay and Lodman, Jackie and Field, Scott E. and Khanna, Gaurav and Scheel, Mark A. and Pfeiffer, Harald P. and Gerosa, Davide and Kidder, Lawrence E.",
    title = "{Eccentric binary black hole surrogate models for the gravitational waveform and remnant properties: comparable mass, nonspinning case}",
    eprint = "2101.11798",
    archivePrefix = "arXiv",
    primaryClass = "gr-qc",
    doi = "10.1103/PhysRevD.103.064022",
    journal = "Phys. Rev. D",
    volume = "103",
    number = "6",
    pages = "064022",
    year = "2021"
}

@article{Harry:2010zz,
    author = "Harry, Gregory M.",
    editor = "Marka, Zsuzsa and Marka, Szabolcs",
    collaboration = "LIGO Scientific",
    title = "{Advanced LIGO: The next generation of gravitational wave detectors}",
    doi = "10.1088/0264-9381/27/8/084006",
    journal = "Class. Quant. Grav.",
    volume = "27",
    pages = "084006",
    year = "2010"
}

@article{VIRGO:2014yos,
    author = "Acernese, F. and others",
    collaboration = "VIRGO",
    title = "{Advanced Virgo: a second-generation interferometric gravitational wave detector}",
    eprint = "1408.3978",
    archivePrefix = "arXiv",
    primaryClass = "gr-qc",
    doi = "10.1088/0264-9381/32/2/024001",
    journal = "Class. Quant. Grav.",
    volume = "32",
    number = "2",
    pages = "024001",
    year = "2015"
}

@article{KAGRA:2020tym,
    author = "Akutsu, T. and others",
    collaboration = "KAGRA",
    title = "{Overview of KAGRA: Detector design and construction history}",
    eprint = "2005.05574",
    archivePrefix = "arXiv",
    primaryClass = "physics.ins-det",
    doi = "10.1093/ptep/ptaa125",
    journal = "PTEP",
    volume = "2021",
    number = "5",
    pages = "05A101",
    year = "2021"
}

@article{LIGOScientific:2021usb,
    author = "Abbott, R. and others",
    collaboration = "LIGO Scientific, VIRGO",
    title = "{GWTC-2.1: Deep extended catalog of compact binary coalescences observed by LIGO and Virgo during the first half of the third observing run}",
    eprint = "2108.01045",
    archivePrefix = "arXiv",
    primaryClass = "gr-qc",
    reportNumber = "LIGO-P2100063",
    doi = "10.1103/PhysRevD.109.022001",
    journal = "Phys. Rev. D",
    volume = "109",
    number = "2",
    pages = "022001",
    year = "2024"
}

@article{LIGOScientific:2021djp,
    author = "Abbott, R. and others",
    collaboration = "KAGRA, VIRGO, LIGO Scientific",
    title = "{GWTC-3: Compact Binary Coalescences Observed by LIGO and Virgo during the Second Part of the Third Observing Run}",
    eprint = "2111.03606",
    archivePrefix = "arXiv",
    primaryClass = "gr-qc",
    reportNumber = "LIGO-P2000318",
    doi = "10.1103/PhysRevX.13.041039",
    journal = "Phys. Rev. X",
    volume = "13",
    number = "4",
    pages = "041039",
    year = "2023"
}

@article{LIGOScientific:2020ibl,
    author = "Abbott, R. and others",
    collaboration = "LIGO Scientific, Virgo",
    title = "{GWTC-2: Compact Binary Coalescences Observed by LIGO and Virgo During the First Half of the Third Observing Run}",
    eprint = "2010.14527",
    archivePrefix = "arXiv",
    primaryClass = "gr-qc",
    reportNumber = "P2000061",
    doi = "10.1103/PhysRevX.11.021053",
    journal = "Phys. Rev. X",
    volume = "11",
    pages = "021053",
    year = "2021"
}

@article{LIGOScientific:2018mvr,
    author = "Abbott, B. P. and others",
    collaboration = "LIGO Scientific, Virgo",
    title = "{GWTC-1: A Gravitational-Wave Transient Catalog of Compact Binary Mergers Observed by LIGO and Virgo during the First and Second Observing Runs}",
    eprint = "1811.12907",
    archivePrefix = "arXiv",
    primaryClass = "astro-ph.HE",
    reportNumber = "LIGO-P1800307",
    doi = "10.1103/PhysRevX.9.031040",
    journal = "Phys. Rev. X",
    volume = "9",
    number = "3",
    pages = "031040",
    year = "2019"
}

@article{Gayathri:2020coq,
    author = "Gayathri, V. and Healy, J. and Lange, J. and O'Brien, B. and Szczepanczyk, M. and Bartos, Imre and Campanelli, M. and Klimenko, S. and Lousto, C. O. and O'Shaughnessy, R.",
    title = "{Eccentricity estimate for black hole mergers with numerical relativity simulations}",
    eprint = "2009.05461",
    archivePrefix = "arXiv",
    primaryClass = "astro-ph.HE",
    doi = "10.1038/s41550-021-01568-w",
    journal = "Nature Astron.",
    volume = "6",
    number = "3",
    pages = "344--349",
    year = "2022"
}

@article{Hinder:2008kv,
    author = "Hinder, Ian and Herrmann, Frank and Laguna, Pablo and Shoemaker, Deirdre",
    title = "{Comparisons of eccentric binary black hole simulations with post-Newtonian models}",
    eprint = "0806.1037",
    archivePrefix = "arXiv",
    primaryClass = "gr-qc",
    reportNumber = "IGC-08-6-1",
    doi = "10.1103/PhysRevD.82.024033",
    journal = "Phys. Rev. D",
    volume = "82",
    pages = "024033",
    year = "2010"
}

@article{Tanay:2016zog,
    author = "Tanay, Sashwat and Haney, Maria and Gopakumar, Achamveedu",
    title = "{Frequency and time domain inspiral templates for comparable mass compact binaries in eccentric orbits}",
    eprint = "1602.03081",
    archivePrefix = "arXiv",
    primaryClass = "gr-qc",
    doi = "10.1103/PhysRevD.93.064031",
    journal = "Phys. Rev. D",
    volume = "93",
    number = "6",
    pages = "064031",
    year = "2016"
}

@article{Gamba:2021gap,
    author = "Gamba, Rossella and Breschi, Matteo and Carullo, Gregorio and Albanesi, Simone and Rettegno, Piero and Bernuzzi, Sebastiano and Nagar, Alessandro",
    title = "{GW190521 as a dynamical capture of two nonspinning black holes}",
    eprint = "2106.05575",
    archivePrefix = "arXiv",
    primaryClass = "gr-qc",
    doi = "10.1038/s41550-022-01813-w",
    journal = "Nature Astron.",
    volume = "7",
    number = "1",
    pages = "11--17",
    year = "2023"
}

@article{Carullo:2023kvj,
    author = "Carullo, Gregorio and Albanesi, Simone and Nagar, Alessandro and Gamba, Rossella and Bernuzzi, Sebastiano and Andrade, Tomas and Trenado, Juan",
    title = "{Unveiling the Merger Structure of Black Hole Binaries in Generic Planar Orbits}",
    eprint = "2309.07228",
    archivePrefix = "arXiv",
    primaryClass = "gr-qc",
    reportNumber = "VIR-0804A-23",
    doi = "10.1103/PhysRevLett.132.101401",
    journal = "Phys. Rev. Lett.",
    volume = "132",
    number = "10",
    pages = "101401",
    year = "2024"
}

@article{Rodriguez:2018pss,
    author = "Rodriguez, Carl L. and Amaro-Seoane, Pau and Chatterjee, Sourav and Kremer, Kyle and Rasio, Frederic A. and Samsing, Johan and Ye, Claire S. and Zevin, Michael",
    title = "{Post-Newtonian Dynamics in Dense Star Clusters: Formation, Masses, and Merger Rates of Highly-Eccentric Black Hole Binaries}",
    eprint = "1811.04926",
    archivePrefix = "arXiv",
    primaryClass = "astro-ph.HE",
    doi = "10.1103/PhysRevD.98.123005",
    journal = "Phys. Rev. D",
    volume = "98",
    number = "12",
    pages = "123005",
    year = "2018"
}

@article{Rodriguez:2017pec,
    author = "Rodriguez, Carl L. and Amaro-Seoane, Pau and Chatterjee, Sourav and Rasio, Frederic A.",
    title = "{Post-Newtonian Dynamics in Dense Star Clusters: Highly-Eccentric, Highly-Spinning, and Repeated Binary Black Hole Mergers}",
    eprint = "1712.04937",
    archivePrefix = "arXiv",
    primaryClass = "astro-ph.HE",
    doi = "10.1103/PhysRevLett.120.151101",
    journal = "Phys. Rev. Lett.",
    volume = "120",
    number = "15",
    pages = "151101",
    year = "2018"
}

@article{Samsing:2017xmd,
    author = "Samsing, Johan",
    title = "{Eccentric Black Hole Mergers Forming in Globular Clusters}",
    eprint = "1711.07452",
    archivePrefix = "arXiv",
    primaryClass = "astro-ph.HE",
    doi = "10.1103/PhysRevD.97.103014",
    journal = "Phys. Rev. D",
    volume = "97",
    number = "10",
    pages = "103014",
    year = "2018"
}

@article{Zevin:2018kzq,
    author = "Zevin, Michael and Samsing, Johan and Rodriguez, Carl and Haster, Carl-Johan and Ramirez-Ruiz, Enrico",
    title = "{Eccentric Black Hole Mergers in Dense Star Clusters: The Role of Binary\textendash{}Binary Encounters}",
    eprint = "1810.00901",
    archivePrefix = "arXiv",
    primaryClass = "astro-ph.HE",
    reportNumber = "LIGO-P1800275",
    doi = "10.3847/1538-4357/aaf6ec",
    journal = "Astrophys. J.",
    volume = "871",
    number = "1",
    pages = "91",
    year = "2019"
}

@article{Zevin:2021rtf,
    author = "Zevin, Michael and Romero-Shaw, Isobel M. and Kremer, Kyle and Thrane, Eric and Lasky, Paul D.",
    title = "{Implications of Eccentric Observations on Binary Black Hole Formation Channels}",
    eprint = "2106.09042",
    archivePrefix = "arXiv",
    primaryClass = "astro-ph.HE",
    doi = "10.3847/2041-8213/ac32dc",
    journal = "Astrophys. J. Lett.",
    volume = "921",
    number = "2",
    pages = "L43",
    year = "2021"
}

@article{Romero-Shaw:2020thy,
    author = "Romero-Shaw, Isobel M. and Lasky, Paul D. and Thrane, Eric and Bustillo, Juan Calderon",
    title = "{GW190521: orbital eccentricity and signatures of dynamical formation in a binary black hole merger signal}",
    eprint = "2009.04771",
    archivePrefix = "arXiv",
    primaryClass = "astro-ph.HE",
    doi = "10.3847/2041-8213/abbe26",
    journal = "Astrophys. J. Lett.",
    volume = "903",
    number = "1",
    pages = "L5",
    year = "2020"
}

@article{Ramos-Buades:2023yhy,
    author = "Ramos-Buades, Antoni and Buonanno, Alessandra and Gair, Jonathan",
    title = "{Bayesian inference of binary black holes with inspiral-merger-ringdown waveforms using two eccentric parameters}",
    eprint = "2309.15528",
    archivePrefix = "arXiv",
    primaryClass = "gr-qc",
    doi = "10.1103/PhysRevD.108.124063",
    journal = "Phys. Rev. D",
    volume = "108",
    number = "12",
    pages = "124063",
    year = "2023"
}

@article{Ma:2019rei,
    author = "Ma, Sizheng and Yunes, Nicolas",
    title = "{Improved Constraints on Modified Gravity with Eccentric Gravitational Waves}",
    eprint = "1908.07089",
    archivePrefix = "arXiv",
    primaryClass = "gr-qc",
    doi = "10.1103/PhysRevD.100.124032",
    journal = "Phys. Rev. D",
    volume = "100",
    number = "12",
    pages = "124032",
    year = "2019"
}

@article{Nagar:2021gss,
    author = "Nagar, Alessandro and Bonino, Alice and Rettegno, Piero",
    title = "{Effective one-body multipolar waveform model for spin-aligned, quasicircular, eccentric, hyperbolic black hole binaries}",
    eprint = "2101.08624",
    archivePrefix = "arXiv",
    primaryClass = "gr-qc",
    doi = "10.1103/PhysRevD.103.104021",
    journal = "Phys. Rev. D",
    volume = "103",
    number = "10",
    pages = "104021",
    year = "2021"
}

@article{Islam:2024tcs,
    author = "Islam, Tousif",
    title = "{Study of eccentric binary black hole mergers using numerical relativity and an inspiral-merger-ringdown model}",
    eprint = "2403.03487",
    archivePrefix = "arXiv",
    primaryClass = "gr-qc",
    month = "3",
    year = "2024"
}

@article{Islam:2024rhm,
    author = "Islam, Tousif",
    title = "{Straightforward mode hierarchy in eccentric binary black hole mergers and associated waveform model}",
    eprint = "2403.15506",
    archivePrefix = "arXiv",
    primaryClass = "astro-ph.HE",
    month = "3",
    year = "2024"
}

@article{Varma:2018mmi,
    author = "Varma, Vijay and Field, Scott E. and Scheel, Mark A. and Blackman, Jonathan and Kidder, Lawrence E. and Pfeiffer, Harald P.",
    title = "{Surrogate model of hybridized numerical relativity binary black hole waveforms}",
    eprint = "1812.07865",
    archivePrefix = "arXiv",
    primaryClass = "gr-qc",
    doi = "10.1103/PhysRevD.99.064045",
    journal = "Phys. Rev. D",
    volume = "99",
    number = "6",
    pages = "064045",
    year = "2019"
}

@article{Islam:2022laz,
    author = "Islam, Tousif and Field, Scott E. and Hughes, Scott A. and Khanna, Gaurav and Varma, Vijay and Giesler, Matthew and Scheel, Mark A. and Kidder, Lawrence E. and Pfeiffer, Harald P.",
    title = "{Surrogate model for gravitational wave signals from nonspinning, comparable-to large-mass-ratio black hole binaries built on black hole perturbation theory waveforms calibrated to numerical relativity}",
    eprint = "2204.01972",
    archivePrefix = "arXiv",
    primaryClass = "gr-qc",
    doi = "10.1103/PhysRevD.106.104025",
    journal = "Phys. Rev. D",
    volume = "106",
    number = "10",
    pages = "104025",
    year = "2022"
}

@article{Estelles:2020twz,
    author = "Estell\'es, H\'ector and Husa, Sascha and Colleoni, Marta and Keitel, David and Mateu-Lucena, Maite and Garc\'\i{}a-Quir\'os, Cecilio and Ramos-Buades, Antoni and Borchers, Angela",
    title = "{Time-domain phenomenological model of gravitational-wave subdominant harmonics for quasicircular nonprecessing binary black hole coalescences}",
    eprint = "2012.11923",
    archivePrefix = "arXiv",
    primaryClass = "gr-qc",
    doi = "10.1103/PhysRevD.105.084039",
    journal = "Phys. Rev. D",
    volume = "105",
    number = "8",
    pages = "084039",
    year = "2022"
}

@article{Khalil:2021txt,
    author = "Khalil, Mohammed and Buonanno, Alessandra and Steinhoff, Jan and Vines, Justin",
    title = "{Radiation-reaction force and multipolar waveforms for eccentric, spin-aligned binaries in the effective-one-body formalism}",
    eprint = "2104.11705",
    archivePrefix = "arXiv",
    primaryClass = "gr-qc",
    doi = "10.1103/PhysRevD.104.024046",
    journal = "Phys. Rev. D",
    volume = "104",
    number = "2",
    pages = "024046",
    year = "2021"
}

@article{Gamba:2024cvy,
    author = "Gamba, Rossella and Chiaramello, Danilo and Neogi, Sayan",
    title = "{Toward efficient effective-one-body models for generic, nonplanar orbits}",
    eprint = "2404.15408",
    archivePrefix = "arXiv",
    primaryClass = "gr-qc",
    doi = "10.1103/PhysRevD.110.024031",
    journal = "Phys. Rev. D",
    volume = "110",
    number = "2",
    pages = "024031",
    year = "2024"
}

@article{Gupte:2024jfe,
    author = "Gupte, Nihar and others",
    title = "{Evidence for eccentricity in the population of binary black holes observed by LIGO-Virgo-KAGRA}",
    eprint = "2404.14286",
    archivePrefix = "arXiv",
    primaryClass = "gr-qc",
    month = "4",
    year = "2024"
}

@article{Cutler:1994ys,
    author = "Cutler, Curt and Flanagan, Eanna E.",
    title = "{Gravitational waves from merging compact binaries: How accurately can one extract the binary's parameters from the inspiral wave form?}",
    eprint = "gr-qc/9402014",
    archivePrefix = "arXiv",
    reportNumber = "GRP-369",
    doi = "10.1103/PhysRevD.49.2658",
    journal = "Phys. Rev. D",
    volume = "49",
    pages = "2658--2697",
    year = "1994"
}

@article{KAGRA:2013rdx,
    author = "Abbott, B. P. and others",
    collaboration = "KAGRA, LIGO Scientific, Virgo",
    title = "{Prospects for observing and localizing gravitational-wave transients with Advanced LIGO, Advanced Virgo and KAGRA}",
    eprint = "1304.0670",
    archivePrefix = "arXiv",
    primaryClass = "gr-qc",
    reportNumber = "LIGO-P1200087, VIR-0288A-12, JGW-P1808427",
    doi = "10.1007/s41114-020-00026-9",
    journal = "Living Rev. Rel.",
    volume = "19",
    pages = "1",
    year = "2016"
}

\end{document}